\documentclass[pra,aps,twocolumn,showpacs,superscriptaddress]{revtex4}
\usepackage{dcolumn}
\usepackage{bm}
\usepackage{amsmath,amssymb}
\usepackage{graphicx}
\usepackage{hyperref}

\newcommand{\ds}[0]{\displaystyle}
\newcommand{\nolfrac}[2]{\genfrac{}{}{0pt}{}{#1}{#2}}

\newcommand{\sqrtfrac}[2]{\frac{#1}{\sqrt{#2}}}

\newcommand{\bra}[1]{\left<#1\right|}

\newcommand{\ket}[1]{\left|#1\right>}

\newcommand{\braket}[2]{\left<\left.#1\right|#2\right>}
\newcommand{\tcoeff}[3]{\left[#1\right]^{#2}_{#3}}

\begin{document}

\title{Computational studies of x-ray scattering from three-dimensionally-aligned asymmetric-top molecules} 

\author{Stefan Pabst}
\affiliation{Argonne National Laboratory, Argonne, Illinois 60439, USA}
\affiliation{Institut f\"ur Theoretische Physik, Universit\"at Erlangen-Nuremberg, D-91058 Erlangen, Germany}

\author{Phay J. Ho}
\affiliation{Argonne National Laboratory, Argonne, Illinois 60439, USA}

\author{Robin Santra}
\thanks{Corresponding author.}
\affiliation{Argonne National Laboratory, Argonne, Illinois 60439, USA}
\affiliation{Department of Physics, University of Chicago, Chicago, Illinois 60637, USA}
\date{\today}

\begin{abstract}
We theoretically and numerically analyze x-ray scattering from asymmetric-top molecules three-dimensionally aligned using elliptically polarized laser light. A rigid-rotor model is assumed. The principal axes of the polarizability tensor are assumed to coincide with the principal axes of the moment of inertia tensor. Several symmetries in the Hamiltonian are identified and exploited to enhance the efficiency of solving the time-dependent Schr\"odinger equation for each rotational state initially populated in a thermal ensemble. Using a phase-retrieval algorithm, the feasibility of structure reconstruction from a quasi-adiabatically-aligned sample is illustrated for the organic molecule naphthalene.  The spatial resolution achievable strongly depends on the laser parameters, the initial rotational temperature, and the x-ray pulse duration. We demonstrate that for a laser peak intensity of 5~TW/cm$^2$, a laser pulse duration of 100~ps, a rotational temperature of 10~mK, and an x-ray pulse duration of 1~ps, the molecular structure may be probed at a resolution of
1 \AA.
\end{abstract} 

\maketitle

\section{Introduction}
\label{sec.intro}

X-ray diffraction is a powerful method for investigating structures of molecules.
X-ray crystallography has become the standard tool for identifying the structure of large molecules and proteins \cite{Drenth-book,Cher-Zewa-ChemPhysChem09,Wats-Cric-Nature53}.
Recent developments in x-ray sources have opened new opportunities \cite{Neutze-etal-Nature00,Spence-Doak-PRL04,Gaffney-Chapman-Science07,Weierstall-etal-ExpFluids08,Huldt-etal-JStrucBio03,Neutze-etal-RadPhysChem04,Miao-etal-Nature99,Miao-etal-AnnRevBiophys04,Shapiro-etal-PNAS05,Spence-etal-UM01} for imaging membrane proteins and other macromolecules that cannot be crystallized \cite{Sayre-StrucChem02}.
One possibility is single-molecule imaging, where one molecule at a time is probed by an intense x-ray pulse \cite{Neutze-etal-Nature00,Gaffney-Chapman-Science07,Miao-etal-AnnRevBiophys04,Huldt-etal-JStrucBio03,Shapiro-etal-PNAS05}, which subjects the molecule to severe damage \cite{Neutze-etal-Nature00,Howells-etal-ELSPEC09}.
%With the dawn of x-ray free electron lasers (XFELs), ultrafast x-ray diffraction with femtosecond pulses provide the necessary intensity for probing single molecules with one shot before destructive effects dominate \cite{Neutze-etal-Nature00,Gaffney-Chapman-Science07,Neutze-etal-RadPhysChem04}. 
A series of diffraction patterns has to be collected and classified according to the molecular orientation to get full structural information from randomly oriented molecules \cite{Huldt-etal-JStrucBio03}.
New iterative phase-retrieval algorithms for noncrystalline specimens have been developed to invert the diffraction data \cite{Miao-etal-JOSA98,Miao-etal-Nature99,Huldt-etal-JStrucBio03,Gaffney-Chapman-Science07,Miao-etal-AnnRevBiophys04,Elser-Milane-ACA08,Elser-JOSA03}. 

Alternative approaches with laser-aligned molecules have been proposed, where an ensemble of molecules, rather than a single molecule, is exposed to x-ray pulses \cite{Spence-Doak-PRL04,Weierstall-etal-ExpFluids08}. In this approach the radiation dose for each molecule is strongly reduced and stays well below the critical value for damage \cite{Spence-Doak-PRL04}.  
Three-dimensional information on the single-molecule structure can be gained from a well-aligned ensemble by capturing many two-dimensional diffraction patterns. The ability to accumulate the signal over a large number of x-ray pulses reduces the radiation dose further. 
Notice that alignment, rather than orientation, of the molecules is expected to be sufficient for the reconstruction of the molecular structure \cite{Elser-Milane-ACA08,Spence-etal-ACA05}. 
Nevertheless, various methods have been proposed to orient molecules \cite{Frie-Hers-JCP99,Frie-Hers-JPCA99} by the linear Stark effect \cite{Sakai-Mine-PRL03}, the AC Stark effect \cite{Chu-PRA08}, or two-color fields \cite{De-etal-PRL09}.

Much theoretical and experimental research has been done on laser-induced
alignment \cite{Stapelfeldt-Seideman-RMP03,Seid-Hami-AAMOP05,Kim-Felker-JCP96,Larsen-etal-JCP99,Larsen-etal-PRL99,Larsen-etal-PRL00,Sakai-etal-JCP98,Arta-Seid-JCP08,Seideman-JCP95,Seideman-JCP99,Friedrich-Herschbach-PRL95,Seideman-JCP01,Poul-Pero-JCP04,Poul-Ejdr-PRA06,Unde-Suss-PRL05,Daem-Guer-PRL05,Lee-Vill-PRL06,Purc-Bark-PRL09,Hami-Seid-PRA05,Bisg-Poul-PRL04,Bisg-Vift-PRA06,Vift-Kuma-PRA09,Vift-Kuma-PRL07,Kuma-Bisg-JCP06,Guer-Rouz-PRA08,Rouz-Guer-PRA08,Holm-Niel-PRL09,Pero-Poul-PRL03,Pero-Poul-PRA04,Seid-PRL99,Dick-Norr-PRL81,Kell-Dion-PRA00,Gers-Aver-PRA08,Dion-Kell-PRA99,Rouz-Guer-PRA06,Rouz-Rena-PRA07,Ho-Star-JCP09,Ho-Mill-JCP09,Peterson-etal-APL08,Ho-Sant-PRA08,Buth-Sant-PRA08-1,Buth-Sant-JCP08,Seid-CPL96,Rena-Hert-PRA05,Zeng-Zhon-LaserPhys09,Leib-Aver-PRL03,Lee-Vill-PRL04}. 
With nonresonant laser frequencies, an induced dipole moment can be created that couples back to the laser electric field and forces the molecule to be aligned. 
Earlier work has focused on one-dimensional alignment of the most polarizable axis by using linearly polarized light \cite{Stapelfeldt-Seideman-RMP03,Seid-Hami-AAMOP05,Rouz-Guer-PRA06,Dion-Kell-PRA99,Gers-Aver-PRA08,Kell-Dion-PRA00,Friedrich-Herschbach-PRL95,Bisg-Poul-PRL04,Seid-PRL99,Pero-Poul-PRL03,Bisg-Vift-PRA06,Kuma-Bisg-JCP06,Hami-Seid-PRA05,Poul-Ejdr-PRA06,Daem-Guer-PRL05,Poul-Pero-JCP04,Seideman-JCP95,Pero-Poul-PRA04,Seideman-JCP01,Ho-Star-JCP09,Ho-Mill-JCP09,Peterson-etal-APL08,Ho-Sant-PRA08,Buth-Sant-PRA08-1,Buth-Sant-JCP08,Seideman-JCP99,Guer-Rouz-PRA08,Rena-Hert-PRA05,Zeng-Zhon-LaserPhys09,Leib-Aver-PRL03,Lee-Vill-PRL04,Seid-CPL96}. 
The demonstration of three-dimensional alignment \cite{Larsen-etal-PRL00,Unde-Suss-PRL05,Lee-Vill-PRL06,Vift-Kuma-PRA09,Vift-Kuma-PRL07,Rouz-Guer-PRA08,Arta-Seid-JCP08}, which requires an asymmetric-top molecule and an elliptically polarized pulse (or two linearly polarized pulses) opens the door for probing the three-dimensional structure.
The Coulomb explosion technique has been exploited to detect three-dimensional alignment \cite{Lee-Vill-PRL06,Larsen-etal-PRL00}, where rotational temperatures down to 1~K have been accomplished \cite{Even-Jort-JCP00}.

Depending on the duration of the laser pulse, $\tau_L$, relative to the rotational period of the molecule, $\tau_\text{rot}$, the alignment 
dynamics can be classified into three distinct regimes.  In the limits of adiabatic ($\tau_L\gg\tau_\text{rot}$) and impulsive 
($\tau_L\ll\tau_\text{rot}$) alignment, the dynamics can be described analytically \cite{Friedrich-Herschbach-PRL95,Seideman-JCP01}. 
Impulsive alignment reveals the quantum mechanical nature of this process by showing alignment revivals after the laser pulse is over \cite{Seid-PRL99,Ho-Mill-JCP09,Poul-Pero-JCP04,Daem-Guer-PRL05,Rena-Hert-PRA05,Zeng-Zhon-LaserPhys09,Leib-Aver-PRL03,Lee-Vill-PRL04}. 
%Much work has been done to enhance and control this feature by changing laser intensity, pulse duration and delay times between pulses \cite{Seid-PRL99,Guer-Rouz-PRA08,Bisg-Vift-PRA06,Poul-Ejdr-PRA06,Zeng-Zhon-LaserPhys09,Lee-Litv-JPB04,Lee-Vill-PRL04,Leib-Aver-PRL90,Rena-Hert.PRA05,Lori-Tehi-PRA08,Torr-deNal-PRA05}.
In the adiabatic limit, the alignment dynamics follow the laser pulse shape \cite{Buth-Sant-JCP08}. 
No analytic solution exists in the intermediate regime ($\tau_L\approx\tau_\text{rot}$), and the time propagation of the molecular ensemble in the presence
of the laser pulse has to be performed numerically \cite{Seideman-JCP01,Vift-Kuma-PRA09}. 
Especially the quantum mechanical description of three-dimensional alignment \cite{Unde-Suss-PRL05,Rouz-Guer-PRA08,Arta-Seid-JCP08} has proved to be numerically expensive \cite{Vift-Kuma-PRA09}.

A general theory of x-ray diffraction from laser-aligned symmetric-top molecules was developed in Ref. \cite{Ho-Sant-PRA08}.  Applications to 
adiabatically aligned molecules may be found in Refs. \cite{Ho-Sant-PRA08,Ho-Star-JCP09}.  Reference \cite{Ho-Mill-JCP09} describes calculations on x-ray scattering from impulsively aligned molecules, exploiting the alignment revivals for probing field-free molecular structure.  
It has been shown that, for a symmetric-top molecule containing a single heavy scatterer, a holographic algorithm can successfully reconstruct 
the molecular structure from an x-ray scattering pattern \cite{Ho-Star-JCP09}. 

In this article, we discuss x-ray scattering from an ensemble of rigid, asymmetric-top molecules aligned three-dimensionally by elliptically polarized light at finite rotational temperature. 
We point out the symmetries in the quantum mechanical theory of three-dimensionally aligned molecules that can be used to significantly reduce the numerical time propagation. 
We restrict our analysis to electronic and vibrational ground-state configurations and neglect deformations.
Our approach allows us to investigate x-ray diffraction from molecules in all three alignment regimes (adiabatic, impulsive, and intermediate).
In order to probe molecular structure of gas-phase molecules by x-ray scattering, the degree of alignment must be rather high for sufficient resolution.
This favors the adiabatic alignment approach \cite{Poul-Ejdr-PRA06} with low-temperature molecules \cite{Kuma-Bisg-JCP06,Buth-Sant-JCP08}.
In Sec. \ref{sec.theory}, we present the theory and inherent symmetries of the Hamiltonian underlying x-ray diffraction from laser-aligned molecules.
Section \ref{sec.methods} focuses on the numerical implementation and computational efficiency, and basic ideas are presented of the phase-retrieval 
algorithm that is used for structure reconstruction.
In Sec. \ref{sec.results}, the three-dimensional alignment dynamics and their impact on the reconstruction are demonstrated using the example of the organic molecule naphthalene.
We conclude with a discussion of the feasibility and limitations of x-ray diffraction from laser-aligned gas-phase molecules.
Atomic units \cite{Drake-book} are employed throughout, unless otherwise noted.

%%%%%%%%%%%%%%%%%%%%%%%%%%%
\section{Theory}
%%%%%%%%%%%%%%%%%%%%%%%%%%%
\label{sec.theory}
The Hamiltonian for x-ray diffraction from laser-aligned molecules is \cite{Ho-Sant-PRA08}
\begin{eqnarray}
  \label{eq.H_tot}
  \hat H_\text{tot}
  &=&
  \hat H_\text{rot}
  +
  \hat H_\text{XEM}
  +
  \hat H_\text{L}(t)
  +  
  \hat H_\text{X},
\end{eqnarray}
where $\hat H_\text{rot}$ describes the field-free rotational motion of a molecule, $\hat H_\text{XEM}$ is the Hamiltonian of the free quantized x-ray fields, and $\hat H_\text{L}(t)$ and $\hat H_\text{X}$ describe the interactions of a molecule with the laser and x-ray field, respectively. 
The electronic and vibrational states of the molecule are omitted, since we assume that the molecule remains in its electronic and vibrational ground state throughout our discussion.
The x-ray field and its interaction with the molecule is described in a quantized manner. The laser field is formulated as a classical field.
The eigenstates of the noninteracting system, i.e., $\hat H_0=\hat H_\text{rot}+\hat H_\text{XEM}$, are
\begin{eqnarray}
  \label{eq.eigenstates_overall}
  \ket{J\tau M;\{n\}}
  &=&
  \ket{J\tau M} \otimes \ket{\{n\}},
\end{eqnarray}
with
\begin{subequations}
\begin{eqnarray}
  \label{eq.eigenenergy_xem}
  \hat H_\text{XEM}\ket{\{n_i\}}
  &=&
  E^X_{\{n_i\}}\ket{\{n_i\}},
\\
	\label{eq.eigenenergy_Hrot}
  \hat H_\text{rot}\ket{J\tau M}
  &=&
  E^\text{rot}_{J\tau}\ket{J\tau M},
\end{eqnarray}
\end{subequations}
where $\ket{J\tau M}$ are the rotational eigenstates of an asymmetric-top molecule \cite{Arta-Seid-JCP08,Zare} and $\ket{\{n\}}$ are the x-ray Fock states  \cite{Ho-Sant-PRA08}.
The density matrix of the whole system is
\begin{eqnarray}
  \label{eq.dens-mat_tot}
  \hat \rho_\text{tot}
  &=&
  \sum_{J\tau M} \sum_{\{n_1\},\{n_2\}} \rho^X_{\{n_1\},\{n_2\}} w_{J\tau}
\\\nonumber&&\times
	\ket{\Psi_{J\tau M;\{n_1\}}(t)} \bra{\Psi_{J\tau M;\{n_2\}}(t)},
\end{eqnarray}
where the gas-phase molecules \cite{Sakurai,Rohringer-Santra-PRA07} are described by a canonical ensemble \cite{Kroto,Reichl-book,Blum-book}, $w_{J\tau}$ is the statistical weight, and $\rho^X_{\{n_1\},\{n_2\}}$ denotes the initial distribution of all the occupied field modes \cite{Loud-book,Mand-Wolf-book}.

With including both interactions, $\hat H_\text{L}(t)$ and $\hat H_\text{X}$, the states of Eq. (\ref{eq.eigenstates_overall}) are no longer eigenstates of the system. 
However, each state $\ket{\Psi_{J_0\tau_0 M_0;\{n_0\}}(t)}$ can be written as
\begin{eqnarray}
	\label{eq.states_coeff}
	\ket{\Psi_{J_0\tau_0 M_0;\{n_0\}}(t)}
	&=&
	\sum_{J_1\tau_1M_1,\{n_1\}}
\\\nonumber&&\times
	\tcoeff{C(t)}{J_1\tau_1M_1;\{n_1\}}{J_0\tau_0 M_0;\{n_0\}}
	\ket{J_1\tau_1M_1;\{n_1\}},
\end{eqnarray}
where the expansion coefficients $\tcoeff{C(t)}{J_1\tau_1M_1;\{n_1\}}{J_0\tau_0 M_0;\{n_0\}}$ statisfy the initial condition 
\begin{eqnarray}
	\label{eq.c_init}
	\tcoeff{C(-\infty)}{J_1\tau_1M_1;\{n_1\}}{J_0\tau_0 M_0;\{n_0\}}
	&=&
	\delta_{J_0J_1}\delta_{\tau_0\tau_1}\delta_{M_0M_1}\delta_{\{n_0\},\{n_1\}}. 
\end{eqnarray}

The equation of motion for the expansion coefficients in the interaction picture (subscript I) reads
\begin{widetext}
\begin{eqnarray}
	\label{eq.eom1}
	i\frac{d}{dt}\tcoeff{C_I(t)}{J_2\tau_2M_2;\{n_2\}}{J_0\tau_0 M_0;\{n_0\}}
	&=&
	\hskip-2ex \sum_{J_1\tau_1M_1,\{n_1\}}\hskip-2ex
	\bra{J_2\tau_2M_2;\{n_2\}}
	\hat H_{L}(t) + \hat H_{X}
	\ket{J_1\tau_1M_1;\{n_1\}}
\\&&\times\nonumber
	e^{i(E^\text{rot}_{J_2\tau_2}-E^\text{rot}_{J_1\tau_1}+E^X_{\{n_2\}}-E^X_{\{n_1\}})t}
	\ \tcoeff{C_I(t)}{J_1\tau_1M_1;\{n_1\}}{J_0\tau_0 M_0;\{n_0\}}.
\end{eqnarray}
\end{widetext}

In the following, we assume we found $\tcoeff{C_{L,I}(t)}{J_1\tau_1M_1}{J_0\tau_0 M_0}$, the solution for the laser-only problem, i.e., $\hat H_\text{rot}+\hat H_\text{L}(t)$. Note that in the laser-only Hamiltonian no x-ray field is involved and we will drop the x-ray field indices in its solution.

The interaction between the laser-aligned molecules and the x-ray field is taken into account by first-order perturbation theory. The solution of Eq. (\ref{eq.eom1}) becomes
\begin{widetext}
\begin{eqnarray}
	\label{eq.eom_solution}
	\tcoeff{C_I(t)}{J_4\tau_4M_4;\{n_4\}}{J_0\tau_0 M_0;\{n_0\}}
	&=&
	-i\hskip-3ex\sum_{\nolfrac{J_1\tau_1M_1, J_2\tau_2M_2}{J_3\tau_3M_3}} \hskip-2ex
	\tcoeff{C_{L,I}(t)}{J_4\tau_4M_4}{J_3\tau_3 M_3}
	\int_{-\infty}^t\!\!dt'\
  e^{i(E^\text{rot}_{J_2\tau_2}-E^\text{rot}_{J_1\tau_1}+E^X_{\{n_4\}}-E^X_{\{n_0\}})t'}
\\&&\times\nonumber
  \tcoeff{C^{-1}_{L,I}(t')}{J_3\tau_3M_3}{J_2\tau_2 M_2}
  \bra{J_2\tau_2M_2;\{n_4\}}
	\hat H_{X}
	\ket{J_1\tau_1M_1;\{n_0\}}
  \tcoeff{C_{L,I}(t')}{J_1\tau_1M_1}{J_0\tau_0 M_0}
  .
\end{eqnarray}
The expectation values of interest can be calculated by
\begin{eqnarray}
  \label{eq.expval}
  {\cal O}(t)
  &=&
  \text{Tr}\left[\hat O \hat\rho_{\text{tot}}(t)\right]
\\\nonumber&=&
  \label{eq.expval_detail}
  \hskip-3ex\sum_{\nolfrac{J_2\tau_2 M_2,J_1\tau_1M_1,J'\tau'M'}{\{n_2\},\{n_1\},\{n'\},\{n''\}}}\hskip-4ex
  w_{J'\tau'}\ \rho^X_{\{n'\},\{n''\}}
  \bra{J_2\tau_2 M_2;\{n_2\}}\hat O\ket{J_1\tau_1M_1;\{n_1\}}
\\\nonumber&&\times
  e^{i(E^\text{rot}_{J_2\tau_2}-E^\text{rot}_{J_1\tau_1}+E^X_{\{n_2\}}-E^X_{\{n_1\}})t}
  \tcoeff{C_I(t)}{J_1\tau_1M_1;\{n_1\}}{J'\tau' M';\{n'\}}
  \tcoeff{C^*_I(t)}{J_2\tau_2 M_2;\{n_2\}}{J'\tau' M';\{n''\}},
\end{eqnarray}
\end{widetext}
from which the alignment signals (cf. Sec. \ref{ssec.cos2}) and the x-ray scattering probability (cf. Sec. \ref{ssec.angu_prob_dist}) can be derived.
We followed here the strategy that is laid out in Ref. \cite{Ho-Sant-PRA08} for x-ray diffraction from symmetric-top molecules and adapt it to the asymmetric-top case.  
The structures of $\hat H_{XEM}$ and $\hat H_X$ are the same for symmetric-top and asymmetric-top molecules.
The most dramatic changes for asymmetric-top molecules happen in the laser-only system.
Therefore, we focus for the rest of our discussion just on the laser-only system.

In the following subsections the structure of the field-free Hamiltonian $\hat H_\text{rot}$ and the laser--molecule interaction matrix $\hat H_{L}(t)$ will be investigated. For this purpose it is convenient to stay in the Schr\"odinger picture. In Sec. \ref{ssec.eom} we return to the equation of motion of the laser-only system and point out the symmetries of its solutions $\tcoeff{C_{L,I}(t)}{J'\tau'M'}{J\tau M}$.
The measure of three-dimensional alignment is described in Sec. \ref{ssec.cos2}.
The theory section closes with exploiting the symmetries in the angular density distribution and the diffraction signal.

\subsection{Free asymmetric-top rotor}
\label{ssec.Hrot}
%%%%%%%%%%%%%%%%%%%%%%%%%%%%%%%%%%%%%%%%%%%%%%
Assuming that structural deformation of the molecule may be neglected,
we treat the molecules as rigid rotors. The corresponding field-free Hamiltonian reads \cite{Zare}
\begin{eqnarray}
	\label{eq.Hrot}
	\hat H_\text{rot}
	&=&
	A\hat J^2_a + B\hat J^2_b + C\hat J^2_c
\\\nonumber
  &=&
	\frac{A + B}{2}\hat{J}^2
	+ \frac{2C-A-B}{2} \hat{J}_0^2
\\\nonumber&&
	+ \frac{A-B}{2}\left[\hat J_{+1}^2+\hat J_{-1}^2\right],
\end{eqnarray}
where $A,B,C$ are the rotational constants associated with the principal axes of inertia,
$\hat J_a,\hat J_b,\hat J_c$ are the Cartesian components of the angular momentum operator in the molecular frame; and $\hat J_{\pm1},\hat J_0$ are the spherical basis components. 
For symmetric-top ($A=B\neq C$) and asymmetric-top ($A\neq B\neq C, A\neq C$) rotors, the angular momentum $J$ and its projection on the space-fixed $z$-axis, $M$, are conserved.  
The angular-momentum projection on the molecular $c$-axis, $K$, is conserved only for symmetric-top molecules.
A new quantum number $\tau$, replacing $K$, must be introduced for asymmetric-top rotors diagonalizing the field-free Hamiltonian [cf. Eq. (\ref{eq.eigenenergy_Hrot})].
Note, the rotational energies $E^\text{rot}_{J\tau}$ are independent of the quantum number $M$. Hence, each energy level $E^\text{rot}_{J\tau}$ is ($2J+1$)-fold degenerate. 
It is possible to express the asymmetric-top eigenstates $\ket{J\tau M}$ as a superposition of the analytically known symmetric-top eigenstates $\ket{JKM}$ \cite{Arta-Seid-JCP08},
\begin{eqnarray}
	\label{eq.eigenstate_asymm}
	\ket{J\tau M}
	&=&
	\sum_K a^{[J]}_{K\tau} \ket{JKM},
\\
	\label{eq.eigenstate_symm}
	\braket{\phi,\theta,\chi}{JKM}
	&=&
	\sqrt{\frac{2J+1}{8\pi^2}}D^{*[J]}_{M,K}(\phi,\theta,\chi),
\end{eqnarray}
where the coefficients $a^{[J]}_{K\tau}$ are real and $\phi,\theta,\chi$ are the three Euler angles connecting the space-fixed laser frame (L) with the principal axes of inertia in the molecular reference frame (M).
$D^{*[J]}_{M,K}(\phi,\theta,\chi)$ is the complex conjugate of the Wigner D-matrix with angular momentum $J$.
Unfortunately, it is not possible to find a general relation between the $\tau$ and $K$ classification \cite{Land-Lifs-book1}.

There are two symmetries incorporated between the symmetric-top ($K$) and asymmetric-top ($\tau$) classification.
The first symmetry decouples states with even $K$ from states with odd $K$, since $\hat H_\text{rot}$ is a linear combination of 
$\hat J^2_{\pm1}$ and $\hat J^2_0$ [Eq. (\ref{eq.Hrot})]. The state class with even $K$ is labeled $E$, the one with odd $K$ is labeled $O$.
Asymmetric eigenstates inherit this separation and are superpositions of just even or odd $K$ states.
The invariance of $\hat H_\text{rot}$ under the substitutions $M\rightarrow -M$ and $K\rightarrow -K$ leads to the second symmetry, the Wang symmetry \cite{Zare,Kroto}, where asymmetric-top eigenstates decompose in symmetric and antisymmetric linear combinations, 
\begin{eqnarray}
	\label{eq.state_pm-K_symm}
	\ket{J\tau M}
	&=&
	\sum_{K\geq0} a^{[J]}_{K\tau}\Big[\ket{JKM}+(-1)^\tau \ket{J(-K)M}\Big]
	,\qquad
\end{eqnarray}
where $(-1)^\tau$ denotes the parity of $\tau$,
\begin{eqnarray}
	\label{eq.asymm_parity}
	(-1)^\tau
	&=&
	\begin{cases}
		+1,&\quad\tau \text{ symmetric in $K$,} \\
		-1,&\quad\tau \text{ antisymmetric in $K$}.
	\end{cases}
\end{eqnarray}
Thus, overall there are four separate state classes, which are summarized in Table \ref{tab.rotational_class}.

\begin{table}[ht!]
	\caption{\label{tab.rotational_class}
		Rotational classes of asymmetric-top rotors.}
	\begin{ruledtabular}
  \begin{tabular}{ll}
	  $E^+:$ &  $\ds \ket{J\tau M} = \sum_{K\geq0,\,K\text{ even}} a^{[J]}_{K\tau}\Big[\ket{JKM}+\ket{J(-K)M}\Big]$   \\
	  $E^-:$ &  $\ds \ket{J\tau M} = \sum_{K\geq0,\,K\text{ even}} a^{[J]}_{K\tau}\Big[\ket{JKM}-\ket{J(-K)M}\Big]$   \\
 	  $O^+:$ &  $\ds \ket{J\tau M} = \sum_{K\geq0,\,K\text{ odd}} a^{[J]}_{K\tau}\Big[\ket{JKM}+\ket{J(-K)M}\Big]$   \\
 	  $O^-:$ &  $\ds \ket{J\tau M} = \sum_{K\geq0,\,K\text{ odd}} a^{[J]}_{K\tau}\Big[\ket{JKM}-\ket{J(-K)M}\Big]$   \\
%	  \rule[0em]{0em}{1.2em}
	\end{tabular}
  \end{ruledtabular}
\end{table}

With knowing the energy levels of $\hat H_\text{rot}$, it is possible to calculate the laser-only density matrix
\begin{eqnarray}
  \label{eq.dens-mat_mol}
  \hat \rho^\text{mol}(t)
  &=&
  \sum_{J\tau M} w_{J\tau} \ket{\Psi_{J\tau M}(t)}\bra{\Psi_{J\tau M}(t)},
\end{eqnarray}
with the statistical weights,
\begin{eqnarray}
	\label{eq.stat_weight}
	w_{J\tau} 
	&=&
	g_{J\tau}\frac{e^{-E^\text{rot}_{J\tau}/kT}}{Z(T)},
\end{eqnarray}
where $Z(T)$ is the partition function at temperature $T$ and $k$ is the Boltzmann constant.
Every state is additionally weighted by the nuclear spin statistical weight $g_{J\tau}$, which represents the number of allowed nuclear spin states for a given rotational state and can be derived from symmetry arguments that have to obey spin statistics \cite{Paul-PR40}.
The computation of $g_{J\tau}$ for asymmetric-top molecules and in particular for naphthalene is discussed in Appendix \ref{app.spin_weight}.

\subsection{Laser--molecule interaction}
\label{ssec.H_L}
%%%%%%%%%%%%%%%%%%%%%%%%%%%%%%%%%%%%%%%%%%%%%%
Three-dimensional alignment may be achieved by using
an intense, nonresonant, elliptically polarized laser field, 
\begin{eqnarray}
	\label{eq.laser_field}
	{\bm E}(t)
	\!\!&=&\!\!
	\sqrt{8\pi\alpha I(t)} \Big(
	  \epsilon_x\cos(\omega t)\,{\bm e}_x 
	  +
  	\epsilon_z\sin(\omega t)\,{\bm e}_z
	\Big),
\end{eqnarray}
where $I(t)$ is the intensity of the laser field, $\alpha$ is the fine structure constant, $\omega$ is the laser frequency, and ${\bm e}_z,{\bm e}_x$ are the unit vectors of the major and minor polarization directions.  The parameters $\epsilon_x$ and $\epsilon_z$ satisfy $\epsilon_x^2+\epsilon_z^2=1$ and $0 \le \epsilon_x \le \epsilon_z$. The laser--molecule interaction reads \cite{Unde-Suss-PRL05}
\begin{eqnarray}
	\label{eq.HL_general}
	\hat H_L(t)
	&=&
	-\frac{1}{2}\sum_{i,j \in \{x,y,z\}}\alpha^\text{pol}_{ij}\,U_{ij}(t)
\\\nonumber
	&=&
	-\frac{1}{2} \sum_{J=0}^2\sum_{M=-J}^J (-1)^{J+M}[\alpha^\text{pol}]^{[J]_L}_M U^{[J]_L}_{-M}(t),
\end{eqnarray}
where $\alpha^\text{pol}$ is the dipole-polarizability tensor and $U(t) = {\bm E}(t)\otimes {\bm E}(t)$ is the electric-field tensor. On the right-hand side of Eq. (\ref{eq.HL_general}), the interaction is written first as a Cartesian tensor product and then as a spherical tensor product. 
Only spherical tensor components with $J=0,2$ are non-zero and contribute to the tensor product, since both tensors are symmetric. (All components with $J=1$ are zero for symmetric tensors.)
The $J=0$ component shifts all rotational energy levels by a state-independent amount and may therefore be dropped.

The laser period, $2\pi/\omega$, is typically several orders of magnitude smaller than the rotational time scale, $\tau_\text{rot} \approx 1/(A+B)$. Cycle averaging leads to a diagonal $U_{ij}$, and only three spherical components remain nonzero,
\begin{subequations}
\label{eq.efield_tensor}
\begin{eqnarray}
	\label{eq.efield_tensor_20}
	[U(t)]^{[2]_L}_{0}
	&=&
	\sqrtfrac{4\pi\alpha}{6}\left(2-3\epsilon_x^2\right) I(t),
\\ 
	\label{eq.efield_tensor_22}
	[U(t)]^{[2]_L}_{\pm2}(t)
	&=&
	2\pi\alpha \epsilon_x^2 I(t).
\end{eqnarray}
\end{subequations}
Hence, only the terms involving $[\alpha^\text{pol}]^{[2]_L}_{0}$ and $[\alpha^\text{pol}]^{[2]_L}_{\pm2}$ contribute to the laser--molecule interaction
[Eq. (\ref{eq.HL_general})]. This holds for any molecule. However, the polarizability $\alpha^\text{pol}$ is a molecular property and is therefore most conveniently expressed in the molecular reference frame $[\alpha^\text{pol}]^{[J]_M}_{K}$. Wigner D-matrices provide the connection to the space-fixed components \cite{Unde-Suss-PRL05},
\begin{eqnarray}
	\label{eq.tensor_trafo}
	[\alpha^\text{pol}]^{[J]_L}_{M}
	&=&
	\sum_{K} D^{*[J]}_{M,K}(\phi,\theta,\chi)\	[\alpha^\text{pol}]^{[J]_M}_{K}.
\end{eqnarray}

When the principal axes of the polarizability tensor do not coincide with the principal axes of the moment-of-inertia tensor, the polarizability tensor is not diagonal in the molecular reference frame (cf. Sec. \ref{ssec.Hrot}) and all $[\alpha^\text{pol}]^{[J]_M}_{K}$ for $J=0,2$ may be nonzero. We will restrict our discussion to the case where both frames coincide. The remaining tensor components are \cite{Unde-Suss-PRL05}
\begin{subequations}
\label{eq.poliby_tensor}
\begin{eqnarray}
	\label{eq.poliby_tensor_20}
	[\alpha^\text{pol}]^{[2]_M}_{0}
	&=&
	\sqrtfrac{2\alpha^\text{pol}_{cc}-\alpha^\text{pol}_{bb}-\alpha^\text{pol}_{aa}}{6},
\\
	\label{eq.poliby_tensor_22}
	[\alpha^\text{pol}]^{[2]_M}_{\pm2}
	&=&
	\frac{\alpha^\text{pol}_{aa}-\alpha^\text{pol}_{bb}}{2}.
\end{eqnarray}
\end{subequations}
Using Eqs. (\ref{eq.efield_tensor}) - (\ref{eq.poliby_tensor}), the matrix elements of the laser--molecule interaction operator with respect to the symmetric-top eigenstates read 
\begin{widetext}
\begin{eqnarray}
	\label{eq.HL_matele}
 \bra{JKM}\hat H_L(t)\ket{J'K'M'}
  &=&
  -\frac{1}{2}\sqrt{\frac{2J+1}{2J'+1}}
\\\nonumber &&\hskip-25ex
  	\times
		\bigg(  
		  [\alpha^\text{pol}]^{[2]_M}_{0}
			\braket{J,K;2,0}{J',K'}
			+[\alpha^\text{pol}]^{[2]_M}_{2} %(\alpha_{aa}-\alpha_{bb})
			\Big[\braket{J,K;2,2}{J'K'}+\braket{J,K;2,-2}{J',K'}\Big]
  	\bigg) 
\\\nonumber && \hskip-25ex
    \times
    \bigg(  	
    	[U(t)]^{[2]_L}_{0}\braket{J,M;2,0}{J',M'}
    	+[U(t)]^{[2]_L}_{2}
  		\Big[\braket{J,M;2,-2}{J',M'}+\braket{J,M;2,2}{J',M'}\Big]
  	\bigg),
\end{eqnarray}
\end{widetext}
where the matrix elements of the Wigner D-matrices have been expressed in terms of Clebsch-Gordan coefficients \cite{Buth-Sant-PRA08-1,Zare}.

Three-dimensional alignment of asymmetric-top molecules conserves neither $K$ nor $M$. The asymmetric-top rotor breaks the $\chi$-symmetry 
($K$ conservation), and elliptically polarized light breaks the $\phi$-symmetry ($M$ conservation).
However, there are remaining symmetries in the laser interaction that can be employed,
\begin{subequations}
\label{eq.HL_symm}
\begin{eqnarray}
	\nonumber	
  \label{eq.HL_evenodd}
  \hskip-4ex
  \bra{JKM}\hat H_L(t)\ket{J'K'M'}
\\&&\hskip-25ex =
  0, 
  \qquad K'-K,M'-M \notin\{\pm2,0\},
\\&&\hskip-25ex =
  \label{eq.HL_pm-M_symm}
	(-1)^{J-J'}
	\bra{JK(-M)}\hat H_L(t)\ket{J'K'(-M')},
\\&&\hskip-25ex =
  \label{eq.HL_pm-K_symm}
	(-1)^{J-J'}
  \bra{J(-K)M}\hat H_L(t)\ket{J'(-K')M'}.
\end{eqnarray}
\end{subequations}
This follows from properties of the Clebsch-Gordan coefficients \cite{Zare}.
As a result of Eq. (\ref{eq.HL_evenodd}), states with even $K$ ($M$) remain separate from states with odd $K$ ($M$).
The consequences of Eqs. (\ref{eq.HL_pm-M_symm}) and (\ref{eq.HL_pm-K_symm}) will be discussed in the following sections.

\subsection{Equation of motion}
\label{ssec.eom}
%%%%%%%%%%%%%%%%%%%%%%%%%%%%%%%%%%%%%%%%%%%%%%
In the laser-only system, the equation of motion [Eq. (\ref{eq.eom1})] for the initially populated state $\ket{JKM}$ reduces to \cite{Unde-Suss-PRL05}
\begin{eqnarray}
	\label{eq.eom_tcoeff}
  i\frac{d}{dt}\tcoeff{C_{L,I}(t)}{J'\tau'M'}{JKM}
  \hskip-1ex&=&\hskip-2ex
  \sum_{J_1\tau_1M_1}
  \bra{J'\tau'M'}\hat H_L(t)\ket{J_1\tau_1M_1}
\\\nonumber&&\times
  e^{i[E^\text{rot}_{J',\tau'}-E^\text{rot}_{J_1,\tau_1}]t}
  \tcoeff{C_{L,I}(t)}{J_1\tau_1M_1}{JKM}
	,
\end{eqnarray}
where the interaction matrix elements are expressed in the symmetric-top basis and the symmetries from Eqs. (\ref{eq.HL_symm}) can be used.

%The symmetries of $\hat H_L(t)$ [cf. Eqs. (\ref{eq.HL_symm})] pass on to $\tcoeff{C_{L,I}(t)}{J'K'M'}{JKM}$ through Eq. (\ref{eq.eom_tcoeff}),
The symmetries of the rotational eigenstates $\ket{J\tau M}$ [cf. Eq. (\ref{eq.state_pm-K_symm}) and Table \ref{tab.rotational_class}] in combination with the symmetries of $\hat H_L(t)$  [cf. Eqs. (\ref{eq.HL_symm})] pass on to $\tcoeff{C_{L,I}(t)}{J'K'M'}{JKM}$ such that
\begin{subequations}
\label{eq.c-symm}
\begin{eqnarray}
	\label{eq.c_M-symm}
	\tcoeff{C_{L,I}(t)}{J'K'M'}{JKM}
	&=&
	(-1)^{J-J'}
	\tcoeff{C_{L,I}(t)}{J'K'(-M')}{JK(-M)},
\\&=&
	\label{eq.c_K-symm_tau}
	(-1)^{J-J'}
	\tcoeff{C_{L,I}(t)}{J'(-K')M'}{J(-K)M}.
\end{eqnarray}
\end{subequations}
% 
%To see this link, it is helpful to assume that all $\tcoeff{C_{L,I}(t_0)}{J'K'M'}{JKM}$ would fulfill Eqs. (\ref{eq.c-symm}) at one point in time $t_0$. If this is true, all $\frac{d}{dt}\tcoeff{C_{L,I}(t_0)}{J'K'M'}{JKM}$ fulfill Eqs. (\ref{eq.c-symm}) as well. Hence, the equation of motion conserves these symmetries and Eqs. (\ref{eq.c-symm}) hold for $t\geq t_0$. Furthermore, with the initial condition, $\tcoeff{C_{L,I}(-\infty)}{J'K'M'}{JKM}=\delta_{JJ'}\delta_{KK'}\delta_{MM'}$ [cf. Eq. (\ref{eq.c_init})], Eqs. (\ref{eq.c-symm}) hold for all $t$.

The laser--molecule interaction preserves the separation of even or odd $K$ and $M$ states [cf. Eq. (\ref{eq.HL_evenodd})] but breaks the Wang symmetry [cf. Eq. (\ref{eq.state_pm-K_symm})]. 
The indices from the asymmetric-top basis transform to the symmetric-top basis according to
\begin{eqnarray}
	\label{eq.c_K-tau_trafo}
	\tcoeff{C_{L,I}(t)}{J'K'M'}{JKM}
	 &=&
%	\sum_{\tau'} a^{[J']}_{\tau'K'} 
%  	\tcoeff{C_{L,I}(t)}{J'\tau'M'}{JKM}
%  =
  \sum_{\tau,\tau'} a^{[J']}_{\tau'K'}a^{[J]}_{\tau K}  
  	\tcoeff{C_{L,I}(t)}{J'\tau'M'}{J\tau M}.
\end{eqnarray}
The symmetric-top representation is especially favorable for the laser--molecules interaction; the field-free propagation is naturally expressed in the asymmetric-top basis.

\subsection{Measure of alignment}
\label{ssec.cos2}
%%%%%%%%%%%%%%%%%%%%%%%%%%%%%%%%%%%%%%%%%%%%%%
The degree of three-dimensional alignment can be characterized in terms of the quantities $\cos^2\theta_{lm}$, where $\theta_{lm}$ is the angle between the space-fixed axis $l$ and the body-fixed axis $m$. The following relations hold among the $\cos^2\theta_{lm}$ \cite{Rouz-Guer-PRA08}:
\begin{eqnarray} 
	\label{eq.cos2_relation}
	\sum_l \cos^2\theta_{lm_0}
	=
	\sum_m \cos^2\theta_{l_0m}
	=
	1
	\quad \forall \, m_0, \, l_0,
\end{eqnarray}
where five of the six relations are independent. These relations reduce the number of independent $\cos^2\theta_{lm}$ to 4.
The matrix elements $\bra{JKM}\cos^2\theta_{lm}\ket{J'K'M'}$ of one set of independent $\cos^2\theta_{lm}$ ($l\in\{x,z\},m\in\{a,c\}$) are given in Appendix \ref{app.alignment}.
The symmetries of $\cos^2\theta_{lm}$ are the same as for $\hat H_L(t)$,
\begin{subequations}
\label{eq.cos2_symm}
\begin{eqnarray}
	\nonumber
  \label{eq.cos2_evenodd}
  \hskip-4ex
  \bra{JKM}\cos^2\theta_{lm}\ket{J'K'M'}
\\&&\hskip-27ex =
  0,
  \qquad K'-K,M'-M \notin\{\pm2,0\},
\\&&\hskip-27ex =
	\label{eq.cos2_pm-M_symm}
	(-1)^{J-J'}
	\bra{JK(-M)}\cos^2\theta_{lm}\ket{J'K'(-M')},
\\&&\hskip-27ex =
	\label{eq.cos2_pm-K_symm}
	(-1)^{J-J'}
	\bra{J(-K)M}\cos^2\theta_{lm}\ket{J'(-K')M'}.
\end{eqnarray}
\end{subequations}
The restriction imposed on $K'-K$ by Eq. (\ref{eq.cos2_evenodd}) makes it attractive to store the matrix elements of $\cos^2\theta_{lm}$ in the symmetric-top basis; the same is true for $\hat H_L(t)$.

The ensemble-averaged expectation values at time $t$ are
\begin{subequations}
\label{eq.cos2_ensemble}
\begin{eqnarray}
	\left<\cos^2\theta_{lm}\right>(t)
	&=&
	\sum_{J\tau M} w_{J\tau}\left<\cos^2\theta_{lm}\right>_{J\tau M}(t),
\\
	\left<\cos^2\theta_{lm}\right>_{J\tau M}(t)
	&=&
	\sum_{J_1K_1M_1}\sum_{J_2K_2M_2}
\\\nonumber&&
  \times
  \bra{J_1K_1M_1}\cos^2\theta_{lm}\ket{J_2K_2M_2}
\\\nonumber&&\times
	\tcoeff{C_L(t)}{J_2K_2M_2}{J\tau M}\tcoeff{C_L^*(t)}{J_1K_1M_1}{J\tau M}
\end{eqnarray}
\end{subequations}
with
\begin{eqnarray}
  \label{eq.tcoeff_s-pic}
  \tcoeff{C_L(t)}{J'K'M'}{J\tau M}
  &=&
  \sum_{\tau'} a^{[J']}_{\tau'K'}
  e^{-iE^\text{rot}_{J',\tau'}t}
  \tcoeff{C_{L,I}(t)}{J'\tau'M'}{J\tau M}.
\end{eqnarray}
Making use of the symmetries in Eqs. (\ref{eq.c-symm}) and (\ref{eq.cos2_symm}), each $\left<\cos^2\theta_{lm}\right>_{JKM}(t)$ fulfills the relations
\begin{subequations}
\label{eq.cos2(t)_symm}
\begin{eqnarray}
	\label{eq.cos2(t)_K_symm}
  \left<\cos^2\theta_{lm}\right>_{JKM}(t)
	&=&
	\left<\cos^2\theta_{lm}\right>_{J(-K)M}(t)
	,
\\
	\label{eq.cos2(t)_M_symm}
	\left<\cos^2\theta_{lm}\right>_{J\tau M}(t)
	&=&
	\left<\cos^2\theta_{lm}\right>_{J\tau (-M)}(t)
	.
\end{eqnarray}
\end{subequations}
Eq. (\ref{eq.cos2(t)_K_symm}) cannot be used for asymmetric-top molecules, since $K$ is not a good quantum number. However, this equation can be exploit for symmetric-top and linear rotor calculations.

\subsection{Angular probability distribution}
\label{ssec.angu_prob_dist}
%%%%%%%%%%%%%%%%%%%%%%%%%%%%%%%%%%%%%%%%%%%%%%
For the calculation of the x-ray scattering probability $\frac{dP}{d\Omega}$ the solution for the full Hamiltonian [Eq. (\ref{eq.H_tot})] has to be known.
The laser--only problem is solved as described in Sec. \ref{ssec.eom}.
The x-ray interaction is treated in first-order perturbation theory as indicated in Eq. (\ref{eq.eom_solution}).
Furthermore, the following assumptions of the x-ray pulse are made: (1) The coherence time of the x-ray pulse is significantly larger as the rotational time scale of the molecules. (2) The bandwidth of the x-ray pulse is much larger than any rotational transition energy. (3) The angular spread of the x-ray pulse is considerably small.
In Ref. \cite{Ho-Sant-PRA08}, the detailed derivation of $\frac{dP}{d\Omega}$ is given with the final result
\begin{eqnarray}
	\label{eq.scatter_prob}
	\frac{dP}{d\Omega}
	&=&
  \frac{d\sigma_\text{th}}{d\Omega}
  S(\bm Q),
\end{eqnarray}
where $d\sigma_\text{th}/d\Omega$ is the Thomson scattering cross section,
\begin{eqnarray}
	\label{eq.diff_signal}
	S(\bm Q)
	&=&
	\int_{-\infty}^\infty\!\!\!dt\ j_X(t)
	\iiint d\phi\, d\theta\, d\chi \sin\theta
\\\nonumber&&
  \times
	\rho(\phi,\theta,\chi;t)\
	\left|F_\text{mol}(\bm Q,\phi,\theta,\chi)\right|^2
\end{eqnarray}
is the diffraction signal, and $j_X(t)$ is the x-ray flux.
The molecular information in the diffraction signal is contained in the molecular form factor,
\begin{eqnarray}
	\label{eq.form-fac}
	F_\text{mol}(\bm Q,\phi,\theta,\chi)
	&=&
	\int d^3r_M\ \rho(\bm r_M) e^{-i{\bm Q}\big[\bm R(\phi,\theta,\chi)\,\bm r_M\big]}
\end{eqnarray}
which is a function of the molecular orientation, i.e., Euler angles, and the momentum transfer $\bm Q$ given in the space-fixed frame (L). The integration $d^3r_M$ is done in the molecular rest frame (M) with $\rho(\bm r_M)$ being the electron density of the molecule.
The rotation matrix $\bm R(\phi,\theta,\chi)$ transforms the vector $\bm r_M$ into the space-fixed frame.
The angular probability distribution,
\begin{eqnarray}
	\label{eq.probability_ensemble}
	\rho(\phi,\theta,\chi;t)
	&=&
	\sum_{J\tau M} 
		w_{J\tau} 
		\left|\braket{\phi,\theta,\chi}{\Psi_{J\tau M}(t)}\right|^2,
\end{eqnarray}
is linked to the Wigner D-matrices $D^{[J]}_{M,K}(\phi,\theta,\chi)$ through Eqs. (\ref{eq.eigenstate_asymm}) and (\ref{eq.eigenstate_symm}).

From the fact that the molecules are aligned rather than oriented, symmetries additional to the ones of $D^{[J]}_{M,K}(\phi,\theta,\chi)$ \cite{Vars-Mosk-book} enter into $\rho(\phi,\theta,\chi;t)$.
These additional symmetries originate from Eqs. (\ref{eq.c-symm}) and the separation of the rotational classes $E^\pm$ and $O^\pm$ throughout the alignment process (cf. Sec. \ref{ssec.H_L}).
In terms of Euler angels, the symmetries of $\rho(\phi,\theta,\chi;t)$ are:
\begin{subequations}
\label{eq.angulardens_symm}
\begin{eqnarray}
	\label{eq.angulardens_symm1}
	\hskip-5ex
	\rho(\phi,\theta,\chi;t)
	\nonumber
\\&&\hskip-5ex =
	\rho(\phi+\pi,\theta,\chi;t)
	=
	\rho(-\phi,\pi-\theta,\chi+\pi;t),
\\&&\hskip-5ex =
	\label{eq.angulardens_symm2}
	\rho(\phi,\theta,\chi+\pi;t)
	=
	\rho(\phi+\pi,\pi-\theta,-\chi;t).
\end{eqnarray}
\end{subequations}
The first two symmetries [Eq. (\ref{eq.angulardens_symm1})] correspond to $C_2$ rotations about the space-fixed axes $z$ and $x$, respectively.
The last two symmetries [Eq. (\ref{eq.angulardens_symm2})] correspond to $C_2$ rotations about the body-fixed axes $c$ and $a$, respectively.
In Eq. (\ref{eq.form-fac}), these symmetries translate into a replacement of $\bm R(\phi,\theta,\chi)$ by $\bm C_{2,x/z}\bm R(\phi,\theta,\chi)$ and $\bm R(\phi,\theta,\chi)\bm C_{2,a/c}$, respectively.
By letting the $\bm C_{2,x/z}$ rotations act on $\bm Q$ rather than on $\bm r_M$, the $S(\bm Q)$ symmetries are found,
\begin{subequations}
\label{eq.diffsig_symm}
\begin{eqnarray}
	\label{eq.diffsig_symm1}
	S(Q_x,Q_y,Q_z)
	&=&
	S(-Q_x,Q_y,Q_z),
\\
	\label{eq.diffsig_symm2}
	&=&
	S(Q_x,-Q_y,Q_z),
\\
	\label{eq.diffsig_symm3}
	&=&
	S(Q_x,Q_y,-Q_z).
\end{eqnarray}
\end{subequations}
Additionally, the Friedel law \cite{Als-Nielsen-McMorrow}, i.e., $F^*(\bm Q) = F(-\bm Q) \Rightarrow S(\bm Q)=S(-\bm Q)$, has been used. 
The symmetries of Eq. (\ref{eq.angulardens_symm2}) can be used to reduce the integration range of the Euler angles in Eq. (\ref{eq.diff_signal}) but only the symmetries of Eq. (\ref{eq.angulardens_symm1}) survive the integration, which are expressed in Cartesian coordinates in Eqs. (\ref{eq.diffsig_symm}).
Regardless of their internal structure, Eqs. (\ref{eq.diffsig_symm}) hold for all aligned molecules. Note that alignment does not distinguish between parallel and anti-parallel orientations.  
Thus, even for perfect alignment, i.e., $\phi=\theta=\chi=0$, four distinct molecular orientations contribute incoherently to the diffraction signal $S(\bm Q)$ \cite{Spence-etal-ACA05}.

%%%%%%%%%%%%%%%%%%%%%%%%%%%
\section{Numerical methods}
\label{sec.methods}
%%%%%%%%%%%%%%%%%%%%%%%%%%%
Earlier work has addressed the problem of numerical efficiency in the computational treatment of three-dimensional alignment of 
asymmetric-top molecules \cite{Vift-Kuma-PRA09}.  
In Sec. \ref{ssec.align_dyn}, we describe numerical techniques and symmetry arguments we have implemented to decrease the numerical effort. 
As far as we are aware, these specific points have not been discussed earlier in the literature. 
In Sec. \ref{ssec.phase-retrieval}, we explain the phase-retrieval method we employ to reconstruct molecular structures from
x-ray scattering patterns.

\subsection{Alignment dynamics}
\label{ssec.align_dyn}
%%%%%%%%%%%%%%%%%%%%%%%%%%%
The density matrix of a canonical ensemble is a sum of density matrices $\ket{J\tau M}w_{J\tau}\bra{J\tau M}$ [cf. Eq. (\ref{eq.dens-mat_mol})]. Analogously, all observables can be written as a sum of independent contributions, which may be calculated separately. 
This was shown for $\left<\cos^2\theta_{lm}\right>$ [Eq. (\ref{eq.cos2_ensemble})] and $\rho(\phi,\theta,\chi;t)$ [Eq. (\ref{eq.probability_ensemble})].
Each contribution has a well-defined behavior under the substitution $M\rightarrow -M$, which can be used to avoid the calculation of $\ket{\Psi_{J\tau M}(t)}$ for $M<0$. 
The same is true for the quantum number $K$, but unfortunately $K$ is not a good quantum number for asymmetric-top molecules, and the wave function $\ket{\Psi_{J\tau M}(t)}$ has to be known to calculate the expectation values.
However, $\ket{\Psi_{J\tau M}(t)}$ can be written as a superposition of $\ket{\Psi_{JKM}(t)}$ [cf. Eq. (\ref{eq.eigenstate_asymm})], and by using the same argument as for $M$, only $\ket{\Psi_{JKM}(t)}$ for $K\geq0$ are necessary to build all $\ket{\Psi_{J\tau M}(t)}$ [cf. Eq.(\ref{eq.state_pm-K_symm})].
This makes it attractive to propagate $\ket{\Psi_{JKM}(t)}$ rather than $\ket{\Psi_{J\tau M}(t)}$.
Taking both symmetries together only the states $\ket{\Psi_{JKM}(t)}$ for $0\leq K,M\leq J$ have to be propagated to understand the full system response, which can save up to a factor 4 in computational effort.

The decoupling between even and odd $K/\tau$-states and $M$-states also enhances efficiency and does not get destroyed by the presence of the interaction $\hat H_L(t)$.
Therefore, many $\tcoeff{C_L(t)}{J'\tau' M'}{JKM}$ remain zero throughout the alignment process.
This holds for symmetric-top as good as for asymmetric-top rotors, since each $\ket{J\tau M}$ can be classified into the class $E^\pm$ or $O^\pm$ (cf. Sec. \ref{ssec.Hrot}).
The benefit is not just a speed-up by a factor 4; also the memory requirement to store $\tcoeff{C_L(t)}{J'\tau' M'}{JKM}$, the matrix $\hat H_L(t)$, and the matrices $\cos^2\theta_{lm}$ is reduced by a factor 4. Memory size can become an issue on PCs when high $J$ states
%, $J\gtrsim 70$,
are not negligible. 

A significant time factor in the calculation, besides the time propagation, is the computation of $\rho(\phi,\theta,\chi;t)$. By employing all symmetries described in Eq. (\ref{eq.angulardens_symm}), the range of Euler angles with nonredundant information is reduced by a factor of 16 compared to the entire domain of Euler angles.

Besides the physical symmetries that can be retrieved from the Hamiltonian, there are two numerical aspects that may improve the propagation speed.
First, the numerical time propagation is commonly done by the fourth-order Runge-Kutta method, where it is important that the change in the wave function per propagation step is small and stays in the convergent region \cite{Mullges-book}. Yet too small step sizes quickly lead to numerical inefficiency. 
For our problem the change in $\tcoeff{C_L(t)}{J'\tau' M'}{JKM}$ is directly proportional to the product $I(t)dt$, where $dt$ is the propagation time step.
With decreasing $I(t)$, $dt$ can be chosen larger without leaving the convergent region.
Variable step sizes are, therefore, important and improve the propagation efficiency further. Numerical tests have shown that the optimized program runs faster by a factor between 3 and 4, where we assumed a Gaussian laser pulse centered at $t_0$ with a full width at half maximum of $\tau_L$.  
Outside of the laser pulse, i.e., $t\notin[t_0-3\tau_L,t_0+3\tau_L]$, the molecules are treated as field-free, and the propagation is performed analytically.

The second improvement takes place at the equation of motion [Eq. (\ref{eq.eom_tcoeff})].
To solve the equation numerically, it has to be discretized in time. 
We chose the fourth-order Runge-Kutta method as our discretization method.
Only the contribution of $\hat H_L(t)$ is approximated when the discretization in time in the equation of motion is made in the interaction picture. The time propagation of $\hat H_\text{rot}$ is analytically exact such that in the field-free case the numerically calculated solution matches the analytical result.
To improve the efficiency in solving the equation of motion, we reduce the need for repeated calculating of $e^{i(E^\text{rot}_{j}-E^\text{rot}_{i})t}$, which goes with $N^2$
for the propagation of all $\tcoeff{C_{L,I}(t)}{J'\tau' M'}{JKM}$, where $N$ is the number of rotational states involved in the propagation.
The evaluation of exponential functions does not occur when the equation of motion is discretized in the Schr\"odinger picture \cite{Seideman-JCP95,Seideman-JCP99,Arta-Seid-JCP08}. 
As a consequence, the propagation of the field-free part is now discretized, which has two major limitations: (1) the field-free propagation is not exact even in the field-free case, where the analytic solution is known, and (2) the propagation step size depends, in addition to the laser--molecule interaction strength, on the highest rotational energy $E^\text{rot}_{J\tau}$. 
In order to use the advantages of 
%the interaction picture and simultaneously improving the efficiency of the propagation by reducing the need of evaluating $e^{i(E^\text{rot}_{j}-E^\text{rot}_{i})t}$
both pictures, we discretize the equation of motion in the interaction picture and transform it afterwards back into the Schr\"odinger picture by reformulating the propagation in terms of $\tcoeff{C_{L}(t)}{J'\tau' M'}{JKM}$. The final discretized equation of motion is:
\begin{eqnarray}
  \label{eq.rk4}
  \tcoeff{C_L(t+dt)}{j}{i}
  &=&
  e^{-2i\varphi_j}
 	\tcoeff{C_L(t)}{j}{i}
\\\nonumber&&\hskip-8ex
 	+
% 	dt\frac{e^{-2i\varphi_j}\tcoeff{D^{(1)}}{j}{i}+2e^{-i\varphi_j}\tcoeff{D^{(2)}}{j}{i}+2e^{-i\varphi_j}\tcoeff{D^{(3)}}{j}{i}+\tcoeff{D^{(4)}}{j}{i}}{6},
	\frac{dt}{6}\Bigg(e^{-2i\varphi_j}\tcoeff{D^{(1)}}{j}{i}+2e^{-i\varphi_j}\tcoeff{D^{(2)}}{j}{i}
\\\nonumber&&
	+2e^{-i\varphi_j}\tcoeff{D^{(3)}}{j}{i}+\tcoeff{D^{(4)}}{j}{i}
	\Bigg)
\end{eqnarray}
\vskip-2ex
\begin{eqnarray}
  \label{eq.rk4-deriv}
  \hskip-3ex
  \tcoeff{D^{(a)}}{j}{i}
  &=&
	-i \sum_{k} [H_L(t)]^j_k \tcoeff{\tilde C_L^{(a)}}{k}{i}
	\ ,a=1,2,3,4,
\end{eqnarray}
where the indices $i,j,k$ are shortcuts for the sets of asymmetric-top quantum numbers, $[H_L(t)]^j_i=\bra{j}\hat H_L(t)\ket{i}$, $e^{-iE^\text{rot}_{j}dt/2}=e^{-i\varphi_j}$, and intermediate solutions are
\begin{subequations}
\label{eq.rk4-inter}
\begin{eqnarray}
	\tcoeff{\tilde C_L^{(1)}}{j}{i}
	&=&
	\tcoeff{C_L(t)}{j}{i},
\\	
	\tcoeff{\tilde C_L^{(2)}}{j}{i}
	&=&
	e^{-i\varphi_j}
  \tcoeff{\tilde C_L^{(1)}}{j}{i}
	+\frac{dt}{2}
	e^{-i\varphi_j}
	\tcoeff{D^{(1)}}{j}{i},
\\
  \tcoeff{\tilde C_L^{(3)}}{j}{i}
	&=&
	e^{-i\varphi_j}
  \tcoeff{\tilde C_L^{(1)}}{j}{i}
  + \frac{dt}{2}
  \tcoeff{D^{(2)}}{j}{i},
\\
  \tcoeff{\tilde C_L^{(4)}}{j}{i}
	&=&
	e^{-2i\varphi_j}
	\tcoeff{\tilde C_L^{(1)}}{j}{i}
  + \frac{dt}{2}
  e^{-i\varphi_j}
 	\tcoeff{D^{(3)}}{j}{i}.
\end{eqnarray}
\end{subequations}
Since $e^{-i\varphi_j}$ depends only on $dt$ and not on the time $t$ itself, it needs to be evaluated only once at the beginning of the propagation. 
During the time propagation, the evaluation of the exponential function is not necessary as long as $dt$ does not change. 
In our simulations, we gained a factor 2 to 5 depending on the simulation parameters, when we employed Eqs. (\ref{eq.rk4})-(\ref{eq.rk4-inter}).
Note, in the field-free limit the time propagation in Eq. (\ref{eq.rk4}) is analytically exact. 

Overall, the use of the symmetries and numerical techniques described allows us to simulate three-dimensional alignment of asymmetric-top molecules from the impulsive to the adiabatic regime up to two orders of magnitude faster.
In the case of quasi-adiabatic alignment of naphthalene at 1~K,
%where 819 states have to be propagated (with exploiting the symmetry arguments disussed), 
the total computation time was 2457~h (41 days) with a PC (CPU: 3~GHz).
The calculation of $S(\bm Q)$ took approximately 8\% of the total time ($\phi,\theta,\chi,t$-grid points: $25\times30\times30\times11$). Without exploiting the symmetries for the Euler angles, this calculation would be almost 1.5 times longer than the time propagation itself.
A detailed list of the physical parameters used for the computations is given in Sec. \ref{sec.results}.

\subsection{Phase-retrieval algorithm}
\label{ssec.phase-retrieval}
%%%%%%%%%%%%%%%%%%%%%%%%%%%%%%%%%%%%%%%
For imperfectly aligned molecules, the diffraction pattern is an incoherent average of single-molecule diffraction patterns of different orientations.  
In the limit that a high degree of molecular alignment is attained, the obtained pattern can be approximated as a single-molecule coherent diffraction 
pattern.  (This is true for naphthalene, since orientation and alignment are equivalent for this molecule.)  One may thus retrieve structural information 
from the single-molecule electron density map, which is the Fourier transform of the single-molecule scattering form factor, $F(\boldsymbol{Q})$.  
Since the diffraction pattern provides only $|F(\boldsymbol{Q})|$, we need to obtain the associated phase before we can recover the molecular structure.
There are iterative numerical algorithms that permit reconstructing the phase directly from the intensity data 
\cite{Gerchberg-Saxton-Opt72, Saxton-Book78, Fienup-AO82, Fienup-etal-JOSA82, Fienup-JOSA87, Miao-etal-JOSA98, Miao-Sayre-ACA00, Marchesini-etal-PRB03, Oszla-Suto-ACA04, Carrozzini-etal-ACA04, Elser-JOSA03, Luke-IP05, Wu-Spence-ACA05, Chapman-etal-JOSA06, Nugent-JOSA07,Marchesini-RSI07, Miao-etal-ARPC08}.  
These algorithms require intensity data sampled at twice the Nyquist frequency.  Successful structural reconstruction using these algorithms has been demonstrated with experimental data \cite{Miao-etal-Nature99, Robinson-etal-PRL01, Miao-etal-PRL02, Weierstall-etal-UM02, Spence-etal-PTRSL02, Nugent-etal-PRL03, Miao-etal-PNAS03, Xiao-Shen-PRB05, Shapiro-etal-PNAS05,Quiney-etal-NP06, Miao-etal-PRL06}.

Here we use the hybrid-input-output (HIO) algorithm \cite{Fienup-AO82, Fienup-etal-JOSA82, Fienup-JOSA87}, which involves iterative Fourier transformation back and forth between the object and Fourier domains.  A solution is found when the known constraints are satisfied in both domains.  We begin by obtaining an initial estimate of the object electronic density via an inverse Fourier transformation of the form factor, $F(\bm Q)=|F(\bm Q)|e^{i \phi_{\bm Q}}$, which is obtained by assigning a random phase, $\phi_{\bm Q}$, to the measured modulus, $|F(\bm Q)|$.  
The random phase $\phi_{\bm Q}$ is chosen such that Friedel's law, $F^*(\bm Q) = F(-\bm Q)$, is satisfied.
With this estimate, we initiate an iterative four-step algorithm, in which the $k$-th iteration is given as follows: \\
1. Fourier transform of the object electron density, $\rho_k(\bm r)$, to obtain $F_{k}(\bm Q)$. \\
2. A Fourier domain operation (FDO) is applied to $F_{k}(\bm Q)$ to obtain $F'_{k}(\bm Q)$ that satisfies the Fourier constraint.  In our FDO, the modulus of $F'_{k}(\bm Q)$ is set to be the measured modulus, $|F(\bm Q)|$, and the phase of $F'_{k}(\bm Q)$ is the  phase of $F_{k}(\bm Q)$. \\
3. Inverse Fourier transform of $F'_{k}(\bm Q)$ to give $\rho'_k(\bm r)$. \\
4. An object domain operation (ODO) is applied to $\rho'_k(\bm r)$ to get a new estimate of the object electron density, $\rho_{k+1}(\bm r)$, that satisfies the object constraint.  Our ODO is given as 
\begin{align}
	&\rho_{k+1}(\bm r)  = 
	\begin{cases} 
 		\rho'_k(\bm r) & \mbox{, for $\bm r\in {\mathcal S}$ and $\rho'_k(\bm r)\geq0$} \\
 		\rho_k(\bm r)-\beta \rho'_k(\bm r) & \mbox{, otherwise,}
	\end{cases}
\end{align}
where $\beta$ is chosen to be 0.9 \cite{Marchesini-etal-PRB03, Chapman-etal-JOSA06, Marchesini-RSI07}, and ${\mathcal S}$ 
is a pre-defined support of the object.

In order to obtain a correctly reconstructed object, a support ${\mathcal S}$ of good quality is needed \cite{Fienup-JOSA87}.  
In our algorithm, the support ${\mathcal S}$ is changed dynamically throughout the HIO algorithm via the Shrink-wrap (SW) 
procedure \cite{Marchesini-etal-PRB03}.  The inverse Fourier transform of the scattering intensity, $|F(\bm Q)|^2$,
equals the auto-correlation function of $\rho(\bm r)$.  Treating the auto-correlation function as a distribution function, 
the initial ${\mathcal S}$ is chosen as a region centered at the mean of the auto-correlation function with a spatial extension 
of two standard deviations.  This choice of ${\mathcal S}$ is fixed during the first 2000 iterations before applying the SW 
procedure periodically after every 200 iterations of the HIO algorithm to obtain a new ${\mathcal S}$.  In the SW procedure, 
the modulus of the object, $|\rho_k(\bm r)|$, is convolved with a Gaussian of width $\sigma$.  The new ${\mathcal S}$ is then 
selected as the region for which the value of the convolved function is above a threshold of 20\% of its maximum \cite{Marchesini-etal-PRB03}.  
The initial width of the Gaussian is chosen to be 2.5 \AA\  and is shrunk linearly to a minimum of 0.25 \AA\ after 400 iterations 
of the SW procedure.  Using the last updated ${\mathcal S}$, an additional 200 iterations of the HIO algorithm are performed.  

\begin{figure}[ht]
  \begin{center}
    \includegraphics[clip,width=\linewidth]{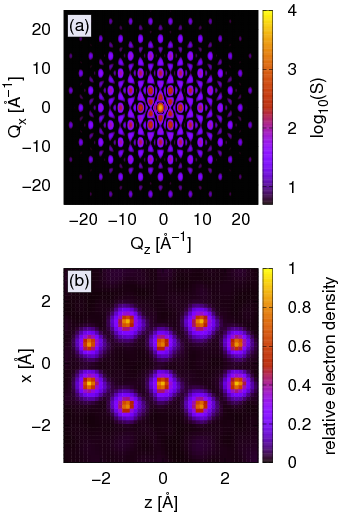}
    \caption{(Color online) (a) The diffraction signal $S(Q_y=0)$ for naphthalene perfectly aligned in the $xz$ plane. 
      The spatial resolution is 0.25~\AA. $S(0)=68^2$ is the number of electrons squared. (b) The reconstructed structure.}
    \label{fig.diff_res_perfect}
  \end{center}
\end{figure}

Figure \ref{fig.diff_res_perfect} illustrates the application of the phase-retrieval algorithm described to an x-ray scattering pattern calculated for perfectly aligned naphthalene molecules with one set of initial random phases. 
X-ray scattering patterns in this article are based on the assumption that the molecular electron
density equals the sum over spherically-averaged atomic electron densities \cite{Als-Nielsen-McMorrow}.  Only the carbon atoms are visible
in the reconstruction in Fig. \ref{fig.diff_res_perfect}(b), since x-ray scattering from carbon is much stronger than x-ray scattering from hydrogen.

Note that the real-space pixel size in each dimension is given by $\pi/Q_\text{max}$, where $Q_\text{max}$ is the maximum momentum transfer
for which diffraction data are available.  In the case of perfect alignment, the real-space resolution, $d_\text{res}$, attained from the 
reconstruction is two times the real-space pixel size.  (In Fig. \ref{fig.diff_res_perfect}, $d_\text{res}=$0.25~\AA.)
However, in the case of imperfect molecular alignment, $d_\text{res}$ depends on $Q_\text{max}$ and $Q_\text{coh}$, where $Q_\text{coh}$ is the 
range of useful diffraction data for which the assumption of coherent scattering holds and the effect of incoherent averaging is small.  In fact, 
we find that including diffraction data beyond $Q_\text{coh}$ can diminish the quality of the reconstructed object.

%%%%%%%%%%%%%%%%%%%%%%%%%%%
\section{Results}
\label{sec.results}
%%%%%%%%%%%%%%%%%%%%%%%%%%%
In this section, we present numerical results based on the theory and the numerical strategies summarized in Secs. \ref{sec.theory} and \ref{sec.methods}.
We demonstrate quasi-adiabatic, three-dimensional alignment of the organic molecule naphthalene ($C_{10}H_8$).
The experimental parameters that we are using are based on previous work \cite{Peterson-etal-APL08}. X-ray energies and fluxes that are used in our discussion are accessible at the Advanced Photon Source at Argonne National Laboratory.

After investigating the alignment dynamics, we focus on the diffraction signal and structure reconstruction. 
The rotational temperature and the x-ray pulse duration impact the effective alignment and limit the structural information stored in the x-ray scattering
pattern.  A phase-retrieval algorithm is used to reconstruct the structure.
In this context, the degree of alignment and its impact on the spatial resolution are discussed.

\subsection{Three-dimensional alignment}
\label{ssec.alignment}
\begin{figure}[ht]
  \begin{center}
    \includegraphics[clip,width=\linewidth]{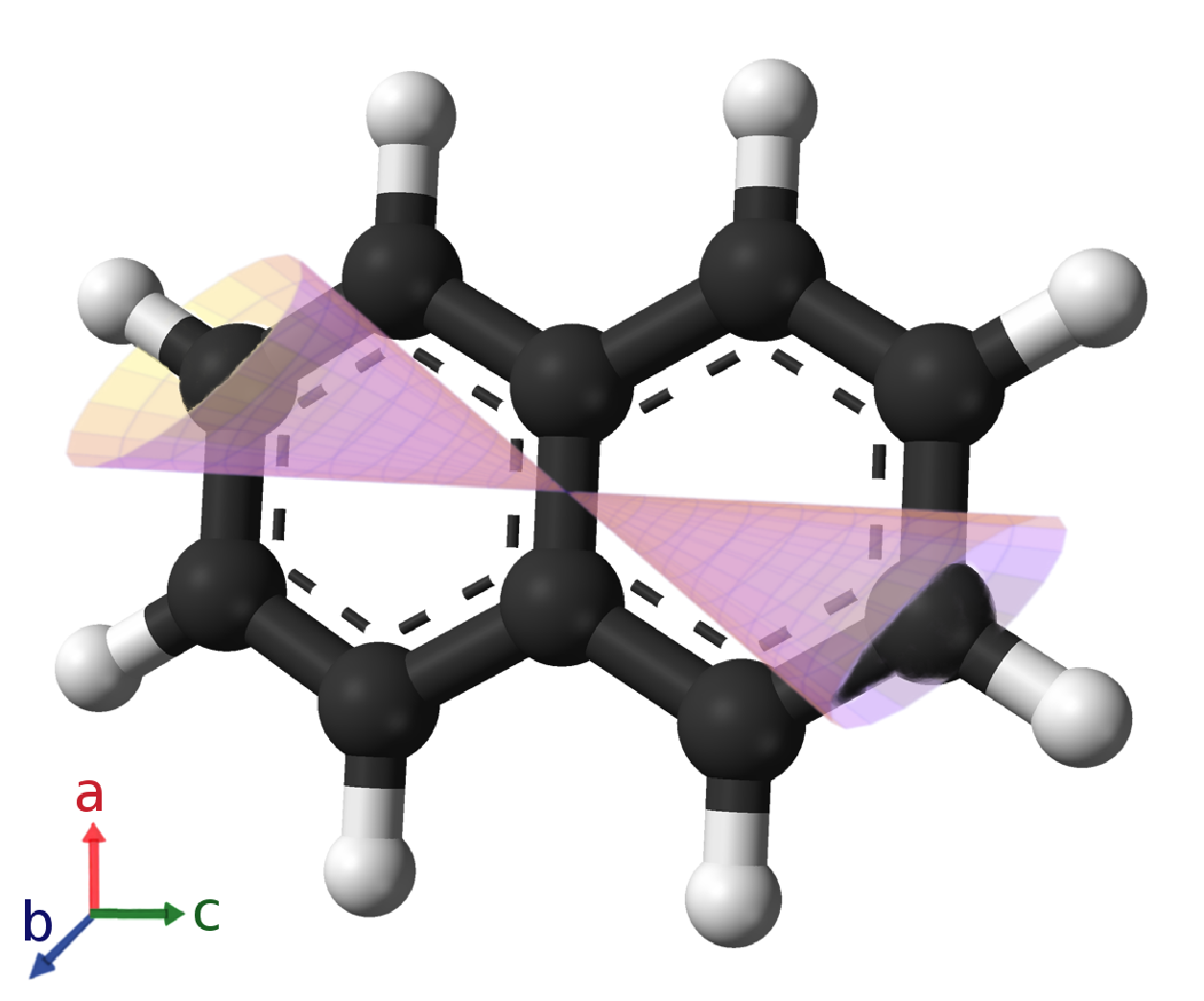}
    \caption{(Color online) The structure of naphthalene ($C_{10}H_8$). All atoms lie in the $ac$ plane of the body-fixed reference frame. 
      The coordinate system is shown in the lower left corner. The cones symbolize the resolution limit of the carbon atoms due to imperfect alignment.}
    \label{fig.naphthalene}
  \end{center}
\end{figure}

%%%%%%%%%%%%%%%%%%%%%%%%%%%
Naphthalene is a planar molecule. 
The molecular reference frame is chosen such that all atoms lie in the $ac$ plane and the $b$ axis is perpendicular to it. 
The choice of axes as well as the structure of naphthalene are shown in Fig. \ref{fig.naphthalene}.
The rotational constants of naphthalene are $A=0.041~\text{cm}^{-1}$, $B=0.029~\text{cm}^{-1}$, and $C=0.104~\text{cm}^{-1}$ \cite{Joo-Taka-JMS02,Hewe-Shen-JCP100}.\footnotetext{In contrast to spectroscopic convention, the rotational constants $A,B$ and $C$ are not ordered according to magnitude. We reorder the rotational constants such that for perfect alignment the molecular frame ($a,b,c$) coincides with the space-fixed frame ($x,y,z$).}
The polarizability constants are $\alpha^\text{pol}_{aa}=121.4~a_0^3$, $\alpha^\text{pol}_{bb}=63.2~a_0^3$, and $\alpha^\text{pol}_{cc}=163.9~a_0^3$ \cite{Howa-Fall-MolPhys99}.
The nuclear spin statistical weights $g_{J\tau}$ for naphthalene are given in Table \ref{tab.spin_weights}. 
(We assume that the carbon nuclei are $^{12}$C isotopes, and the hydrogen nuclei are protons.)
The derivation of $g_{J\tau}$ is outlined in Appendix \ref{app.spin_weight}. 

\begin{table}
	\caption{\label{tab.spin_weights}
		The nuclear spin statistical weights $g_{J\tau}$ of naphthalene classified by the rotational symmetry classes.}
	\begin{ruledtabular}
	\begin{tabular}{c|cccc}
		Rotation class & E$^+$ & E$^-$ & O$^+$ & O$^-$ \\\hline
		$g_{J\tau}$ ($J$ even)& 76 & 60 & 60 & 60 \\
		$g_{J\tau}$ ($J$ odd) & 60 & 76 & 60 & 60 \\
	\end{tabular}
	\end{ruledtabular}
\end{table}

Three-dimensional alignment of naphthalene is obtained by employing elliptically polarized light.  The $z$ and $x$ axes of the space-fixed frame
are defined by the major and minor axes, respectively, of the ellipse that characterizes the elliptically polarized light.
Following Ref. \cite{Rouz-Guer-PRA08}, we choose the ratio between the field components using 
\begin{eqnarray}
  \label{eq.bestratio}
  \frac{\epsilon_x^2}{\epsilon_z^2}
  &=&
  \frac{\alpha^\text{pol}_{cc}-\alpha^\text{pol}_{aa}}
    {\alpha^\text{pol}_{cc}-\alpha^\text{pol}_{bb}}
    = 0.422,
\end{eqnarray}
which maximizes the three-dimensional alignment of naphthalene.
Perfect three-dimensional alignment with respect to the space-fixed frame is achieved, 
when $\left<\cos^2\theta_{xa}\right>=\left<\cos^2\theta_{yb}\right>=\left<\cos^2\theta_{zc}\right>=1$. Random orientation corresponds
to $\left<\cos^2\theta_{lm}\right>=1/3$ for all angles.

The typical distance between two neighboring carbon atoms in naphthalene is 1.4~\AA. In order to resolve the atomic structure, the resolution 
must be smaller than this value.  Imperfect alignment limits the resolution.
To build a connection between the resolution and the alignment in terms of $\left<\cos^2\theta_{lm}\right>$, it is helpful to view the 
angle $\theta_{lm}$ as an opening angle within which the residual motion of the atoms takes place (cf. Fig. \ref{fig.naphthalene}). 
Hence, the length characterizing the smallest resolvable structure is 
\begin{eqnarray}
  \label{eq.align_res}
  d_\text{coh} 
  & \approx &
  2R \sqrt{1-\left<\cos^2\theta_{lm}\right>},
\end{eqnarray}
where $R$ is the linear dimension of the molecule (measured from the center of mass of the molecule).
The degree of alignment needed for a resolution of 1~\AA\ for the outermost carbon atom ($R=2.5$~\AA) is $\left<\cos^2\theta_{lm}\right>=0.96$.
For a given resolution, smaller opening angles $\theta_{lm}$ are required when the molecules become larger. 
In the adiabatic regime at sufficiently low temperature the maximum alignment for a linear rotor ($A=B,C=0$) is given by 
$\left<\cos^2\theta_{zc}\right>=1-\sqrt{4B/\gamma}, \gamma=\sqrt{96}\pi\alpha I[\alpha^\text{pol}]^{[2]}_0$ 
\cite{Friedrich-Herschbach-PRL95,Seideman-JCP01}. 
The polarizability $[\alpha^\text{pol}]^{[2]}_0$ is approximately proportional to $R^3$, and $B$ scales approximately as $R^{-5}$ \cite{Spence-etal-ACA05,Ho-Mill-JCP09}. 
The smallest resolvable dimension $d_\text{coh}$ is thus proportional to $1/R$ and might be expected to decrease with increasing molecular size.  
Note, however, that the temperature required to suppress thermal effects ($kT/B\ll 1$) also decreases as a function of the molecular size.  
Additionally, in order to remain in the adiabatic regime, the laser pulse duration must increase with increasing molecular size.  
In practice this means the laser intensity will decrease.
For large molecules it is more realistic to consider the high temperature limit ($kT/B\gg 1$), where the degree of alignment is a competitive interplay between rotational temperature and coupling strength $\gamma$; more precisely, $\left<\cos^2\theta_{zc}\right>=1-\sqrt{\pi kT/\gamma}$ \cite{Seideman-JCP01,Kuma-Bisg-JCP06}. By using the same scaling arguments, we find $d_\text{coh}\propto \sqrt[4]{kTR}$. The expected resolution is now reversed and increases with molecular size and rotational temperature. Consequently, larger molecules can be resolved less precisely. 

\begin{figure}[ht]
  \begin{center}
    \includegraphics[clip,width=\linewidth]{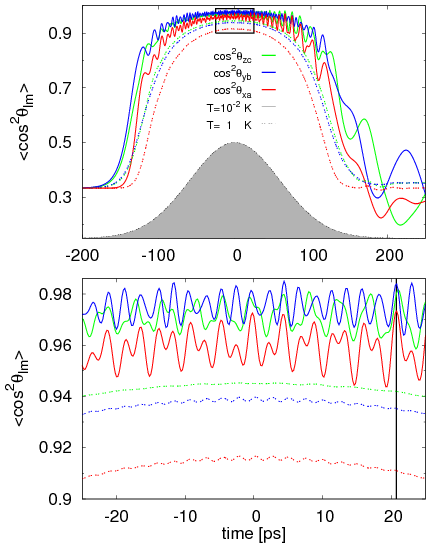}
    \caption{(Color online) Alignment dynamics of three axes of naphthalene at rotational temperatures $T=10$~mK (solid) and $T=1$~K (dotted).
    The pump laser shown in gray has a peak intensity of $I=5$~TW/cm$^2$, a width of $\tau_L=100$~ps, and an ellipticity of $\epsilon^2_x/\epsilon^2_z=0.422$. The upper panel pictures the alignment dynamics on the time scale of the pulse duration. 
    The lower panel is a close-up view of the region highlighted in the upper graph.  
    At $t=20.75$~ps, nonadiabatic oscillation enhances the alignment for all axes at $T=10$~mK and is ideal for a 1-ps x-ray probe pulse.}
    \label{fig.cos2_T10mK}
  \end{center}
\end{figure}

In Fig. \ref{fig.cos2_T10mK}, the alignment dynamics of naphthalene at 10~mK and 1~K, respectively, are shown. We assumed a Gaussian laser pulse with a peak intensity of
$I=5$~TW/cm$^2$ and a pulse duration of $\tau_L=100$~ps (FWHM). 
At a rotational temperature of $T=10$~mK, almost all naphthalene molecules are in the rotational ground state.  The rotational period of naphthalene at such a low temperature, $\tau_\text{rot} \approx 1/(A+B) \approx 476$~ps, is of the order of $\tau_L=100$~ps, suggesting that the alignment dynamics are quasi-adiabatic \cite{Buth-Sant-JCP08}.  This is consistent with the observation that the overall alignment follows the laser pulse shape and, in addition, clear nonadiabatic features (fast oscillations) are visible.  One can make use of the nonadiabatic behavior by probing the molecules with an x-ray pulse of 1 ps duration, which is fast enough to resolve the nonadiabatic oscillations.  For instance, at $t=20.75$~ps the alignment is transiently enhanced in all three dimensions (cf. Fig. \ref{fig.cos2_T10mK}).  
At $T=1$~K, the rotational period of naphthalene is small in comparison to $\tau_L$, so the alignment dynamics are adiabatic and fast oscillations are significantly suppressed.  
As a consequence of the increased thermal motion, the maximum degree of alignment at $T=1$~K is clearly reduced.

\begin{figure}[ht]
  \begin{center}
    \includegraphics[clip,width=\linewidth]{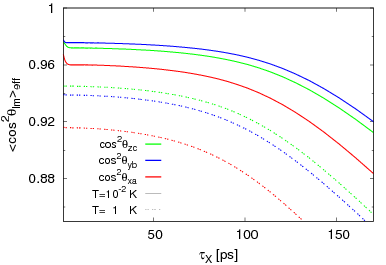}
    \caption{(Color online) The effective alignment of the body-fixed axes is shown as a function of the x-ray pulse duration, $\tau_X$, 
      for $T=10$~mK (solid) and $T=1$~K (dotted). 
      The pump laser has a peak intensity of $I=5$~TW/cm$^2$, a width of $\tau_L=100$~ps, and an ellipticity of $\epsilon^2_x/\epsilon^2_z=0.422$.
      The x-ray pulses are centered at $t=0$~ps.}
    \label{fig.cos2_eff}
  \end{center}
\end{figure}

The diffraction signal $S(\bm Q)$ [cf. Eq. (\ref{eq.diff_signal})] collected over the x-ray pulse duration reflects a pulse-averaged, effective alignment
\begin{eqnarray}
	\left<\cos^2\theta_{lm}\right>_\text{eff}
	&=&
	\int dt\, \bar{j}_X(t)\left<\cos^2\theta_{lm}\right>(t),
\end{eqnarray}
where $\bar{j}_X(t)$ is the normalized x-ray flux.  We assume a Gaussian temporal envelope for the x-ray pulse, with a full width at half maximum of
$\tau_X$.  In Fig. \ref{fig.cos2_eff}, $\left<\cos^2\theta_{lm}\right>_\text{eff}$ is shown as a function of $\tau_X$. 
We may conclude from Fig. \ref{fig.cos2_eff} that the effective alignment of the body-fixed axes decreases rapidly when the x-ray pulse duration is longer than the laser pulse duration.
An enhancement in the effective alignment is visible for $\tau_X\approx1$~ps at $T=10$~mK, where nonadiabatic oscillations are not suppressed and can be resolved by the x-ray pulse.  

\begin{figure}[ht]
  \begin{center}
    \includegraphics[clip,width=\linewidth]{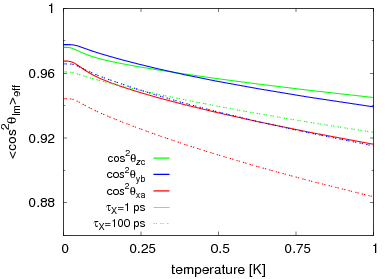}
    \caption{(Color online) The effective alignment of the body-fixed axes is shown as a function of the rotational temperature
      for $\tau_X=1$~ps (solid) and $\tau_X=100$~ps (dotted). The pump laser has a peak intensity of $I=5$~TW/cm$^2$, a width of $\tau_L=100$~ps, and an ellipticity of $\epsilon^2_x/\epsilon^2_z=0.422$. The x-ray pulses are centered at $t=0$~ps.}
   \label{fig.cos2_temp}
  \end{center} 
\end{figure}

When $\tau_X\lesssim \tau_L$, the influence of $\tau_X$ is rather weak and the degree of alignment is mainly affected by the rotational temperature, $T$. 
Figure \ref{fig.cos2_temp} shows the effective alignment of the molecular axes as a function of $T$ for $\tau_X=1$~ps and $\tau_X=100$~ps.
Below 0.25 K, the best aligned axis is the body-fixed $b$ axis in the space-fixed $y$ direction for $\tau_X\leq100$~ps, despite the fact that no laser field is applied in 
this direction. The strong alignment of the $b$ axis translates to well-aligned naphthalene molecules in the polarization plane of the laser
($xz$ plane).  Within the plane, the body-fixed $c$ axis is always more strongly aligned in the $z$ direction than the $a$ axis is aligned 
in the $x$ direction.  (Recall that $\alpha^\text{pol}_{cc}>\alpha^\text{pol}_{aa}$ and $\epsilon_z>\epsilon_x$.)  
For higher temperatures, the body-fixed $c$ axis is the most strongly aligned axis, since the alignment of the $b$ axis is affected by the more rapidly decreasing alignment of the $a$ axis.

As may be seen in Fig. \ref{fig.cos2_eff}, the effective alignment $\left<\cos^2\theta\right>_\text{eff}$ of all molecular axes 
is at least 0.96 for 10~mK naphthalene molecules probed by x-ray pulses shorter than $\sim 20$~ps.
Hence, the smallest resolvable dimension limited by residual pendular motion is $d_\text{coh}=1$~\AA. 
Only the rotational ground state is initially occupied, and the minimum intensity required to accomplish 1~\AA\ resolution is $I=5$~TW/cm$^2$.  Dissociation and ionization play only a minor
role at this intensity, but become important at higher intensities \cite{Smit-Ledi-RCMS98}. 
With an effective alignment of $\left<\cos^2\theta_{xa}\right>_\text{eff}=0.88$ at 1~K and $\tau_X=100$~ps, the smallest resolvable dimension
$d_\text{coh}$ is 1.73~\AA, which is larger than the distance between neighboring carbon atoms.

\subsection{X-ray diffraction patterns}
\label{ssec.diff-pic}
%%%%%%%%%%%%%%%%%%%%%%%%%%%

\begin{figure}[ht]
  \begin{center}
    \includegraphics[clip,width=\linewidth]{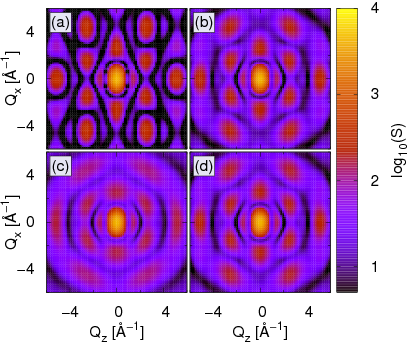}
    \caption{(Color online) X-ray diffraction signal $S(Q_y=0)$ of naphthalene. (a) Perfect alignment.  (b) $T=10$~mK, $\tau_X=1$~ps,
      x-ray pulse centered at $t=20.75$~ps (cf. Fig. \ref{fig.cos2_T10mK}).  
      (c) $T=1$~K, $\tau_x=100$~ps, x-ray pulse centered at $t=0$~ps.
      (d) $T=10$~mK, $\tau_X=100$~ps, x-ray pulse centered at $t=0$~ps.
      For the alignment pulses in (b), (c), and (d), we assumed $I=5$~TW/cm$^2$ and $\tau_L=100$~ps.} 
    \label{fig.diffpic_front}
  \end{center}
\end{figure}

In Fig. \ref{fig.diffpic_front}, diffraction patterns $S(\bm Q)$ of three-dimensionally aligned naphthalene are shown for perfect alignment as well as for imperfect alignment at different temperatures and x-ray pulse widths. The laser parameters are the same as for 
Fig. \ref{fig.cos2_T10mK}. The laser polarization plane is assumed to be the $xz$ plane so that the molecules are aligned as illustrated in Fig. \ref{fig.naphthalene}.  The diffraction patterns in Fig. \ref{fig.diffpic_front} are two-dimensional planar slices through
the three-dimensional $\bm Q$ space, taken at $Q_y=0$.  (Due to the curvature of the Ewald sphere \cite{Als-Nielsen-McMorrow}, experimental 
scattering patterns do not correspond to exactly planar slices through $\bm Q$ space, but this is of no consequence here.) 
The signal strength falls rapidly for high $Q$.  Therefore, to highlight the structure at high momentum transfers, the scattering intensities 
are shown on a logarithmic scale.  The effective alignment in Fig. \ref{fig.diffpic_front} decreases clockwise, from 
$\left<\cos^2\theta_{xa}\right>_\text{eff}=1$ for perfect alignment to $\left<\cos^2\theta_{xa}\right>_\text{eff}=0.88$ for naphthalene molecules at
1~K probed by a 100-ps x-ray pulse.  By increasing the x-ray pulse width from 1 to 100~ps, the effective alignment at $T=10$~mK decreases from $\left<\cos^2\theta_{xa}\right>_\text{eff}=0.967$ to $\left<\cos^2\theta_{xa}\right>_\text{eff}=0.944$ [cf. Fig. \ref{fig.diffpic_front}(b) and (d)], and is significantly smaller than the impact of temperature rise from 10~mK to 1~K [cf. Fig. \ref{fig.diffpic_front}(c) and (d)].

Although the basic features of the x-ray scattering pattern for perfectly aligned naphthalene (Fig. \ref{fig.diffpic_front}a) are preserved in the
scattering patterns for laser-aligned naphthalene [Figs. \ref{fig.diffpic_front}(b), (c), and (d)], the contours are washed out and the contrast between 
maxima and minima is less pronounced with decreasing alignment.  Incoherent averaging for laser-aligned naphthalene renders the diffraction patterns 
more cylindrically symmetric with respect to the $y$ direction, which limits the accessible structural information particularly at high momentum transfer.

\begin{figure}[ht]
  \begin{center}
    \includegraphics[clip,width=\linewidth]{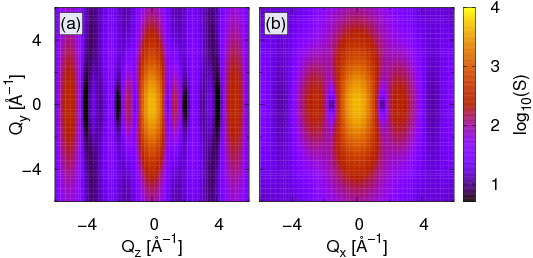}
    \caption{(Color online) (a) Diffraction signal of naphthalene for $Q_x=0$.  (b) Diffraction signal of naphthalene for $Q_z=0$.
      Other parameters are the same as for Fig. \ref{fig.diffpic_front}(b).}
    \label{fig.diffpic_side}
  \end{center}
\end{figure}

Side views of naphthalene in $\bm Q$ space are shown in Fig. \ref{fig.diffpic_side}, where Fig. \ref{fig.diffpic_side}(a) depicts 
the long side ($Q_x=0$) and Fig. \ref{fig.diffpic_side}(b) the short side ($Q_z=0$) of naphthalene.
Both display strong similarities to multi-slit diffraction patterns, consistent with the planar structure of naphthalene being well aligned in the $xz$ plane.

\begin{figure}[ht]
  \begin{center}
    \includegraphics[clip,width=\linewidth]{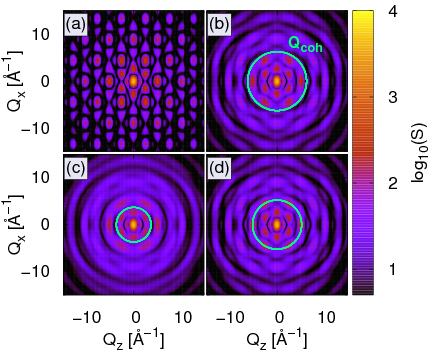}
    \caption{(Color online) Same as for Fig. \ref{fig.diffpic_front}, except $Q_\text{max}=8\pi$~\AA$^{-1}$.  
      The momentum transfer $Q_\text{coh}$ (see the text) is highlighted by a solid green circle in (b)--(d).}
    \label{fig.diffpic_highQ}
  \end{center}
\end{figure}

The achievable real-space resolution depends on two aspects. On the one hand, the smallest resolvable dimension $d_\text{coh}$ is a function of the degree of alignment [Eq. (\ref{eq.align_res})].  On the other hand, the pixel size of the real-space structure reconstruction 
is determined by the maximum momentum transfer $Q_\text{max}$ in the diffraction signal $S(\bm Q)$.  Here, $Q_\text{max}=2\pi$~\AA$^{-1}$, 
corresponding to a pixel size of 0.5~\AA.  
The approximate range in momentum space within which the assumption of coherent scattering holds may be defined by $Q_\text{coh}=2\pi/d_\text{coh}$.  In Fig. \ref{fig.diffpic_highQ}, the diffraction signals of Fig. \ref{fig.diffpic_front} are 
shown up to $Q_\text{max}=8\pi$~\AA$^{-1}$.  The respective ranges defined by $Q_\text{coh}$ are highlighted.  It is not possible to increase the resolution of the real-space structure reconstruction by choosing $Q_\text{max}$ much greater than $Q_\text{coh}$.  In fact, numerical tests have indicated that data beyond $Q_\text{coh}$ can lead to poor convergent structures.

For 12-keV x-ray photons, a real-space pixel size of 0.5~\AA\ requires the detection of photons scattered up to $60^\circ$ with respect to the x-ray propagation axis.  The azimuthally averaged differential x-ray scattering cross section per naphthalene molecule for $Q=2\pi$~\AA$^{-1}$ is $d\sigma/d\Omega_\text{avg}=0.6~\text{barn}$.  Since the largest distance between carbon atoms in naphthalene is $\sim 5$~\AA, the area of a pixel in momentum space must not exceed $4\pi^2/25$~\AA$^{-2}$.  At a scattering angle of $60^\circ$, this corresponds to a solid angle $d\Omega=0.012$.  For a molecular beam width of 1~mm and an x-ray focus area of $100$~$\mu$m$^2$, it has been demonstrated that the number of molecules in the interaction volume can be as large as $10^7$ \cite{Peterson-etal-APL08}.  Hence, requiring a minimum of five scattered photons per pixel \cite{Shen-Baza-JSyncRad04}, the estimated acquisition time for one diffraction pattern is around 70~seconds with an x-ray fluence of $10^{13}$ photons/pulse/mm$^2$ and a repetition rate of 1~kHz.

\subsection{Structure reconstruction}
\label{ssec.recon-pic}
%%%%%%%%%%%%%%%%%%%%%%%%%%%

We applied the phase-retrieval algorithm described in Sec. \ref{ssec.phase-retrieval} to the naphthalene diffraction signals shown in Fig. \ref{fig.diffpic_front}.
The maximum momentum transfer of $2\pi~$\AA$^{-1}$ in Fig. \ref{fig.diffpic_front}
corresponds to a pixel size of 0.5~\AA, which is of the order of the structure we want to resolve in naphthalene. 
As a consequence, each pixel encodes a lot of structure information.
Calculations have shown that the reconstruction routine becomes sensitive to the set of initial random phases. 
Therefore, we follow the spirit of Ref. \cite{Miao-etal-ARPC08} and average over reconstructions obtained for 100 different sets of initial phases to define a quality criterion, which we apply subsequently to all 100 reconstructions to select the most meaningful results and average over these selected reconstructions.
%To define a quality criterion, we assume the unfiltered average over 100 reconstructions is close to the structure we want to recover. 
Our quality criterion is defined as follows: 
(1) All pixels of the unfiltered averaged result that have at least 20\% of the maximum electron density are selected to build a density core region.
(2) The quality criterion uses the density core region and selects only the reconstructions that have at least 50\% of their total electron density within this density core region. 
%Assigning the thresholds to 20\% and 50\%, respectively, in the described criterion is one possible parameter choice; other threshold values may put stricter or looser restrictions on the selected reconstructed structures.

\begin{figure}[ht]
  \begin{center}
    \includegraphics[clip,width=\linewidth]{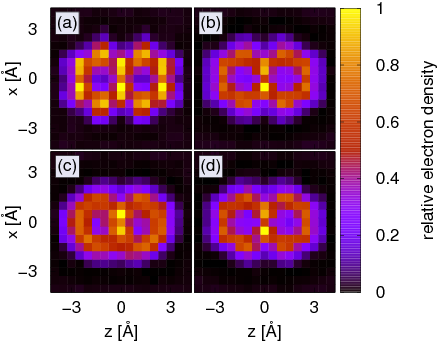}
    \caption{(Color online) Structure of naphthalene, reconstructed from the diffraction signals shown in Fig. \ref{fig.diffpic_front}. 
      The pixel size is 0.5~\AA.
      (a) Perfect alignment.  (b) $T=10$~mK, $\tau_X=1$~ps, x-ray pulse centered at $t=20.75$~ps.
      (c) $T=1$~K, $\tau_X=100$~ps, x-ray pulse centered at $t=0$~ps.
      (d) $T=10$~mK, $\tau_X=100$~ps, x-ray pulse centered at $t=0$~ps.}
    \label{fig.reconstruct_front}
  \end{center}
\end{figure}

The filtered averages for the x-ray scattering patterns of Fig. \ref{fig.diffpic_front} are displayed in Fig. \ref{fig.reconstruct_front}.
In our calculations, we employed a grid spacing in $\bm Q$ space of 0.16~\AA$^{-1}$, corresponding to a maximum object size 8 times larger than the size of naphthalene.

Figure \ref{fig.reconstruct_front} illustrates that with better effective alignment in the diffraction pattern more structural information can be reconstructed. 
In the perfect alignment case (cf. Fig. \ref{fig.reconstruct_front}(a), the positions of the atoms can be resolved.
A comparison with the perfect alignment reconstruction in Fig. \ref{fig.diff_res_perfect}(b), where the pixel size is 0.125~\AA~ and no averaging over several initial random phases was performed, illustrates that the reconstruction of the carbon positions is more difficult when the pixel size is comparable with the size of the atoms. 
Incoherent averaging for imperfectly aligned molecules affects the effective resolution further, since the assumption of coherent scattering limits the usable diffraction signal to $Q\leq Q_\text{coh}$.
%In contrast, information from $Q\geq Q_\text{coh}$ (cf. Fig. \ref{fig.diffpic_highQ}) even destabilizes the reconstruction routine.
As a consequence, $Q_\text{max}$ should be of the order of $Q_\text{coh}$ and, therefore, the pixel size in the reconstruction is limited by the degree of effective alignment.

All results displayed in Fig. \ref{fig.reconstruct_front} recover the two-ring structure of naphthalene, indicating an effective resolution below 2~\AA.
Especially the two carbon atoms connecting both carbon rings are distinctly visible.
Figure \ref{fig.reconstruct_front}(b) and \ref{fig.reconstruct_front}(d) show the reconstruction for naphthalene at 10~mK with $\tau_X=1~$ps and $\tau_X=100~$ps, respectively.
Both have a similar effective alignment and reconstruction.
In both cases, the electron density peaks on the two-ring structure indicate the positions of all carbon atoms and are consistent with the theoretically expected effective resolutions, which are below the typical distance between neighboring carbon atoms (1.4~\AA). 
The predicted effective resolution at 1~K ($d_\text{coh}=1.73$~\AA) is sufficient to resolve the overall structure of naphthalene, which can be seen in Fig. \ref{fig.reconstruct_front}(c). For the central region of naphthalene, the effective resolution improves and makes it possible to identify the two central carbon atoms at 1~K.

%%%%%%%%%%%%%%%%%%%%%%%%%%%
\section{Conclusion}
%%%%%%%%%%%%%%%%%%%%%%%%%%%
\label{sec.conclusion}
We have theoretically studied the theory of x-ray diffraction from asymmetric-top molecules that are three-dimensionally aligned by elliptically polarized light and have discussed a phase-retrieval method in order to reconstruct molecular structure from the x-ray scattering pattern.
The interaction of the gas phase molecules with the laser and x-ray fields was studied in the density matrix formalism. 
We have assumed rigid rotor molecules.
Symmetries in the time-dependent Hamiltonian and in its solutions as well as in the angular density distribution were exploited and used to improve numerical efficiency. 
In combination with further computational aspects, a significant enhancement in numerical efficiency has been achieved. 

A feasibility study of x-ray diffraction from three-dimensionally laser-aligned molecules has been performed using the organic molecule naphthalene.
We have linked the degree of alignment to an effective resolution and have studied the impact of x-ray pulse duration and temperature on the diffraction patterns.
A phase-retrieval reconstruction was performed on diffraction patterns taken for different effective alignments.
The enhancement of incoherent averaging due to imperfect alignment destroys structural information and hinders the recovery of detailed atomic configurations within the molecule.
The reconstruction of naphthalene confirms our discussion that the degree of alignment is a good indicator for the achievable resolution in the reconstruction. 
To decode structures on an atomic length scale, high degree of alignment in all three molecular axes must be achieved.
Molecules have to be cooled down to a few Kelvin or even sub-Kelvin temperatures so rotational motion is sufficiently reduced.
Experiments have shown that it is feasible to cool molecules to 1~K \cite{Even-Jort-JCP00,Fils-Kuep-JCP09,Stap-EuroPhysJD}.
To accomplish rotational temperatures well below 1~K is a great challenge and would open the opportunity to image large gas-phase molecules at atomic resolution.
The problem of incoherent averaging over a finite range of different orientations has to be further addressed, especially in the case where molecular symmetries do not coincide with the symmetries of the diffraction pattern imposed by alignment.

With the ability to detect molecular structure, x-ray scattering from gas phase molecules can be used to study torsion effects and laser-induced deformations that are expected to occur during alignment in the presence of intense laser pulses \cite{Vill-Asey-PRL00}.
By systematically varying the delay time between pump (laser) and probe (x-ray) pulse, it is possible to follow molecular motion on an ultrafast time scale.
Of particular interest is the study of physical and chemical processes in the presence of intense laser fields, which simultaneously provide the required alignment for imaging of reactions with atomic resolution in space and time.

\acknowledgments
We thank Cassandra Hunt for helpful comments on the manuscript.
%We thank .... for discussions.
%S. P. would like to acknowledge gratefully the support of Prof. P.-G. Reinhard.
This work was supported by the Office of Basic Energy Sciences, U.S. Department of Energy under Contract No. DE-AC02-06CH11357.

\appendix
\section{Nuclear spin statistical weights}
\label{app.spin_weight}
%%%%%%%%%%%%%%%%%%%%%%%%%%%
In the density matrix, statistical weights, $w_{J\tau}=g_{J\tau}\frac{\exp(-E^\text{rot}_{J\tau}/kT)}{Z(T)}$, define the relative number of particles in an ensemble that are in a given quantum mechanical state for a well-defined temperature $T$. 
The nuclear spin statistical weights $g_{J\tau}$ are based on symmetry arguments and represent the number of the allowed nuclear spin states for a given rotational state $\ket{J\tau M}$.

The overall symmetry $\Gamma_\text{tot}$ of the total molecular wave function is independent of rotational or nuclear spin states and determined by the spin statistic theorem \cite{Land-Lifs-book1}. However, $\Gamma_\text{tot}$ is also a direct product of  symmetries of the different quantum states \cite{Kroto},
\begin{eqnarray}
  \label{eq.mol_symm}
  \Gamma_\text{tot}
  &=&
  \Gamma_r
  \otimes
  \Gamma_\text{ns},
\end{eqnarray}
where $\Gamma_r$ is the rotational state symmetry and $\Gamma_\text{ns}$ is the nuclear spin state symmetry.
Molecules are in their electronic and vibrational ground states and their contributions can be omitted in Eq. (\ref{eq.mol_symm}).

The representation of symmetries depends on the symmetry group of the molecule.
The molecular symmetry group of asymmetric-top molecules without an inversion center is $D_{2h}$ or lower \cite{Herzberg-book2}. 
The rotational symmetry group of an asymmetric-top rotor is always $V$ (isomorphic to $D_{2h}$) \cite{Herzberg-book2,Herzberg-book3,Bunker-book}.
If there is no common symmetry class between $\Gamma_r$ and $\Gamma_\text{ns}$, $g_{J\tau}=1$.
In other words, molecular symmetries come only into play when molecular rotations are identical to particle exchanges. 

All symmetry representations are written as linear combinations of irreducible representations (irrep) of $V$ in Table \ref{tab.character_table}. 
In addition the character table for the nuclear spin and rotational states are shown. 
The four molecular rotations and particle exchanges defining $V$ are: $E,C_a,C_b,C_c$.

\begin{table}[ht]
	\caption{\label{tab.character_table}
		Character table and irreducible representations of overall, rotational and nuclear spin states for naphthalene.
		The symmetry point group is $V$.}
	\begin{ruledtabular}
  \begin{tabular}{cccccp{1ex}cc}
		&\multicolumn{4}{c}{Operations $P$ of $V$} && \multicolumn{2}{c}{irrep.}\\\cline{2-5}\cline{7-8}
    & $E$ & $C_c$ & $C_b$ & $C_a$ && $J$ even & $J$ odd \rule[-.5em]{0em}{1.5em}\\\hline
    $\Gamma_\text{tot}$ &  1 &  1 &  1 &  1 &&  \multicolumn{2}{c}{$A$} \rule[-.5em]{0em}{1.5em}\\\hline
	  $E^+$      &  1 &  1 & $(-1)^J$     & $(-1)^J$     &&  $A$    &  $B_c$ \\
	  $E^-$      &  1 &  1 & $(-1)^{J+1}$ & $(-1)^{J+1}$ && $B_c$   &   $A$  \\
	  $O^-$      &  1 & -1 & $(-1)^J$     & $(-1)^{J+1}$ && $B_b$   &  $B_a$ \\
	  $O^+$      &  1 & -1 & $(-1)^{J+1}$ & $(-1)^J$     && $B_a$   &  $B_b$ \\\hline
	  $\Gamma_\text{ns}$ & $2^8$ & $2^4$ & $2^4$ & $2^4$ && \multicolumn{2}{l}{$76 A + 60(B_a+B_b+B_c)$}
	  \rule[0em]{0em}{1.2em}
	\end{tabular}
  \end{ruledtabular}
\end{table}

%The characters of each state correspond to the sign changes in the corresponding wave function under the symmetry operations of $V$.
The characters of the overall wave function $\chi_\text{tot}$ can be derived from the Pauli principle \cite{Paul-PR40} and correspond to the overall sign changes induced by the molecular rotations. First, we assume all carbon atoms in naphthalene are $^{12}$C with nuclear spin 0. By remembering only odd permutations of half-integer particles (fermions) change the sign of the overall wave function \cite{Kroto}, we find all group operation leave the overall sign for naphthalene unchanged, since all Hydrogen permutations are even. 
Hence, $\chi_\text{tot}[P]=1,\ \forall P\in V$ and $\Gamma_\text{tot}=A$.
%Therefore, no operation of $V$ does change the sign of the total wave function due to the carbon atoms. Only odd permutations of Hydrogen atoms, which are fermions, can change the sign in naphthalene. However, all Hydrogen permutations are even and do not change the overall sign as well. Hence, $\chi_g[P]=1,\ \forall P\in V$ and $\Gamma_g=A$.

The rotational symmetry classes, in which each rotational state can be classified (cf. Sec. \ref{ssec.Hrot}), coincide with the irreps of $V$ \cite{Zare}. However, whether or not $J$ is even or odd defines which irrep corresponds to which symmetry class (cf. Table \ref{tab.character_table}).
 
The sum of characters over all possible nuclear spin states $\chi_\text{ns}[P]$ for each operation $P$ can be derived by the formula \cite{Land-Lifs-book1,Bunker-book}
\begin{eqnarray}
  \label{eq.spin_character}
  \chi_\text{ns}[P]
  &=&
  \prod_{G_P}\, (2S_{G_P}+1)\ \text{sgn}(P_{G_P})^{2S_{G_P}},
\end{eqnarray}
where the atom group $G_P$ is defined by the minimum number of identical atoms in the molecule such that after the atom permutation $P$ each atom in $G_P$ is replaced by an atom of $G_P$. 
$\text{sgn}(P_{G_P})$ is the sign of the permutation $P$ within in the atom group $G_P$, which is $+1$ or $-1$ depending whether the permutation is even or odd in $G_P$.
The smallest atom group consist of one atom, which does not change position under the operation $P$. 
$S_{G_P}$ is the nuclear spin of the atoms in $G_P$.

Imposing Eq. (\ref{eq.mol_symm}), nuclear spin and rotational symmetries are linked together. (For naphthalene: $\Gamma_\text{tot}=\Gamma_r \otimes\Gamma_\text{ns} =A$).
The multiplication of symmetry representations is done by multiplying their characters.
The nuclear spin statistical weights $g_{J\tau}$ are the number of irreps in $\Gamma_\text{ns}$ such that the product with a given rotational class is $\Gamma_\text{tot}$.
As an example:  When $\ket{J\tau M}\in O^-$ and J is even, then $\Gamma_r \in B_b$ and only the product $B_b\otimes B_b=A$ \cite{Bunker-book} fulfills Eq. (\ref{eq.mol_symm}). Since there are 60 $B_b$ irreps in $\Gamma_\text{ns}$, the nuclear spin statistical weight is 60.
%Here, we dropped the quantum number $M$ in the rotational states, since its classification does not depend on $M$.
For all rotational states of naphthalene, the nuclear spin statistical weights are
\begin{eqnarray}
  \label{eq.spin_weights_naphthalene}
  g_{J\tau}
  &=&
  \begin{cases}
    76 &,~\Gamma_r\in A \\
    60 &\text{, otherwise}.
  \end{cases}
\end{eqnarray}

\section{Matrix elements of $\cos^2\theta_{lm}$}
\label{app.alignment}
%%%%%%%%%%%%%%%%%%%%%%%%%%%
The squares of the matrix elements of the rotation matrix $R(\phi,\theta,\chi)$, i.e. $\cos^2\theta_{lm}$, give a complete picture of the three-dimensional alignment.
Knowing four $\cos^2\theta_{lm}$ is sufficient to describe all, where at least two axes of each reference frame have to be involved in the four $\cos^2\theta_{lm}$. 
Our choice is $l\in\{x,z\},m\in\{a,c\}$, since it is one of the least computationally expensive choices.
The $\cos^2\theta_{lm}$-matrices are diagonal in $M$ or $K$ when $l=z$ or $m=c$, respectively. 
Only the matrix $\cos^2\theta_{xa}$ is nondiagonal in $K$ and $M$.
Each $\cos^2\theta_{lm}$ can be written in terms of $D^{[J]}_{MK}(\phi,\theta,\chi)$ \cite{Seid-CPL96}, where the matrix elements of the Wigner D-functions are given in Refs. \cite{Buth-Sant-PRA08-1,Zare,Gers-Aver-PRA08}.
The matrix elements of the four matrices that we have chosen are:
\begin{widetext}
\begin{subequations}
\begin{eqnarray}
  \label{eq.cos2zc}
  \bra{JKM}\cos^2\theta_{zc}\ket{J'K'M'}
  &=&
  \frac{1}{3} \delta_{JJ'}\delta_{KK'}\delta_{MM'}
  +
  \frac{2}{3}\sqrt{\frac{2J+1}{2J'+1}}
\\\nonumber&& \times
    \braket{J,M;2,0}{J',M'}\braket{J,K;2,0}{J',K'},
\\
  \bra{JKM}\cos^2\theta_{xc}\ket{J'K'M'}
  &=&
  \frac{1}{3} \delta_{JJ'}\delta_{KK'}\delta_{MM'}
  -
  \sqrt{\frac{1}{6}}\sqrt{\frac{2J+1}{2J'+1}} \braket{J,K;2,0}{J',K'} 
\\\nonumber&& \times
  \Bigg(
		\sqrt{\frac{2}{3}}\braket{J,M;2,0}{J',M'}
    -
	  \Big[\braket{J,M;2,2}{J',M'}+\braket{J,M;2,-2}{J',M'}\Big]
  \Bigg),
\\
\bra{JKM}\cos^2\theta_{za}\ket{J'K'M'}
  &=&
  \frac{1}{3} \delta_{JJ'}\delta_{KK'}\delta_{MM'}
  -
  \sqrt{\frac{1}{6}}\sqrt{\frac{2J+1}{2J'+1}}\braket{JM;20}{J'M'}
\\\nonumber&& \times
  \Bigg(
		\sqrt{\frac{2}{3}}\braket{J,K;2,0}{J',K'}
    -
	  \Big[\braket{J,K;2,2}{J',K'}+\braket{J,K;2,-2}{J',K'}\Big]
	\Bigg),
\\
\bra{JKM}\cos^2\theta_{xa}\ket{J'K'M'}
  &=&
  \frac{1}{3} \delta_{JJ'}\delta_{KK'}\delta_{MM'}
  +
  \frac{1}{4}\sqrt{\frac{2J+1}{2J'+1}}
%  \Bigg[\frac{2}{3}\delta_{KK'}\delta_{MM'}
\\\nonumber&& \times
    \Bigg(
    	\sqrt{\frac{2}{3}}\braket{J,M;2,0}{J',M'}
      -
      \Big[\braket{J,M;2,2}{J',M'}+\braket{J,M;2,-2}{J',M'}\Big]
    \Bigg)
\\\nonumber&& \times
    \Bigg(
      \sqrt{\frac{2}{3}}\braket{J,K;2,0}{J',K'}
      -    
      \Big[\braket{J,K;2,2}{J',K'}+\braket{J,K;2,-2}{J',K'}\Big]
    \Bigg).
%\\\nonumber&&\hskip0mm
%   	-\sqrt{\frac{2}{3}}
%    \Big[
%      \braket{J,M;2,2}{J',M'}+\braket{J,M;2,-2}{J',M'}
%    \Big]
%\\\nonumber&&\hskip10mm
%		\times\Big[
% 		\delta_{KK'}\braket{J,K;2,0}{J',K}+\delta_{MM'}\braket{J,M;2,0}{J',M}
%  	\Big]
%\\\nonumber&&\hskip0mm
%    +\Big[\braket{J,K;2,2}{J',K'}+\braket{J,K;2,-2}{J',K'}\Big]
%\\\nonumber&&\hskip10mm
%   \times\Big[\braket{J,M;2,2}{J',M'}+\braket{J,M;2,-2}{J',M'}\Big]
%  \Bigg].
\end{eqnarray}
\end{subequations}
\end{widetext}


\begin{thebibliography}{999}

\bibitem{Wats-Cric-Nature53} J. D. Watson and F. H. C. Crick, Nature {\bf 171}, 737-738 (1953).

\bibitem{Cher-Zewa-ChemPhysChem09} M. Chergui and A. H. Zewail, ChemPhysChem {\bf 10}, 28-43 (2009).

\bibitem{Drenth-book} J. Drenth, \textit{Principles of Protein X-Ray Crystallography} 3rd ed. (Springer, 2006), ISBN 978-0387333342.

\bibitem{Spence-Doak-PRL04} J. C. H. Spence and R. B. Doak, Phys. Rev. Lett. {\bf 92}, 198102 (2004).

\bibitem{Weierstall-etal-ExpFluids08} U. Weierstall, R. B. Doak, J. C. H. Spence, D. Starodub, D. Shapiro, P. Kennedy, J. Warner, G. G. Hembree, P. Fromme, H. N. Chapman, Experiments in Fluids {\bf 44}, 675-689 (2008).

\bibitem{Neutze-etal-RadPhysChem04} R. Neutze, G. Huldt, J. Hajdu, and D. van der Spoel, Radiation Physics and Chemistry {\bf 71}, 905 (2004).

% 7
\bibitem{Shapiro-etal-PNAS05} D. Shapiro, P. Thibault, T. Beetz, V. Elser, M. Howells, C. Jacobsen, J. Kirz, E. Lima, H. Miao, A. M. Neiman, and D. Sayre, Proc. Natl. Acad. Sci. U.S.A. {\bf 102}, 15 343 (2005).

\bibitem{Neutze-etal-Nature00} R. Neutze, R. Wouts, D. van der Spoel, E. Weckert, and J. Hajdu, Nature (London) {\bf 406}, 752 (2000).

\bibitem{Huldt-etal-JStrucBio03} G. Huldt, A. Szoke, and J. Hajdu, Journal of Structural Biology {\bf 144}, 219-227 (2003).

\bibitem{Gaffney-Chapman-Science07} K. J. Gaffney and H. N. Chapman, Science {\bf 316} 1444 (2007).

\bibitem{Miao-etal-AnnRevBiophys04} J. W. Miao, H. N. Chapman, J. Kirz, D. Sayre, and K. O. Hodgson, Annual Review of Biophysics and Biomolecular Structure, {\bf 33}, 157-176 (2004).

% 11
\bibitem{Miao-etal-Nature99} J. Miao, C. Charalambous, J. Kirz, and D. Sayre, Nature (London) {\bf 400}, 342 (1999).

\bibitem{Spence-etal-UM01} J. C. H. Spence, M. R. Howells, L. D. Marks, and J. Miao, Ultramicroscopy {\bf 90}, 1 (2001).

\bibitem{Sayre-StrucChem02} D. Sayre, Structural Chemistry {\bf 13} 81 (2002)

\bibitem{Howells-etal-ELSPEC09} M. R. Howells, T. Beetz, H. N. Chapman, C. Cui, J. M. Holton, C. J. Jacobsen, J. Kirz, E. Lima, S. Marchesini, H. Miao, D. Sayre, D. A. Shapiro, J. C. H. Spence, and D. Starodub, Journal of Electron Spectroscopy and Related Phenomena {\bf 170}, 4-12 (2009).

\bibitem{Elser-Milane-ACA08} V. Elser and R.P. Milane, Acta Cryst. A {\bf 64}, 273 (2008).

%16
\bibitem{Miao-etal-JOSA98} J. Miao, D. Sayre, and H. N. Chapman, J. Opt. Soc. Am. A {\bf 15}, 1662 (1998).

\bibitem{Elser-JOSA03} V. Elser, J. Opt. Soc. Am. A {\bf 20}, 40 (2003).

\bibitem{Spence-etal-ACA05} J. C. H. Spence, K. Schmidt, J. S. Wu, G. Hembree, U. Weierstall, B. Doak and P. Fromme, Acta Cryst. A {\bf 61}, 237 (2005).

\bibitem{Frie-Hers-JCP99} B. Friedrich and D. Herschbach, J. Chem. Phys. {\bf 111}, 6157-6160 (1999).

\bibitem{Frie-Hers-JPCA99} B. Friedrich and D. Herschbach, J. Phys. Chem. A {\bf 103}, 10280 (1999).

%21
\bibitem{Sakai-Mine-PRL03} H. Sakai, S. Minemoto, H. Nanjo, H. Tanji, and T. Suzuki, Phys. Rev. Lett. {\bf 90}, 083001 (2003).

\bibitem{Chu-PRA08} X. Chu, Phys. Rev. A {\bf 78}, 043408 (2008).

\bibitem{De-etal-PRL09} S. De, I. Znakovskaya, D. Ray, F. Anis, N. G. Johnson, I. A. Bocharova, M. Magrakvelidze, B. D. Esry, C. L. Cocke, I. V. Litvinyuk, and M. F. Kling, Phys. Rev. Lett. {\bf 103}, 153002 (2009).

\bibitem{Kuma-Bisg-JCP06} V. Kumarappan, C. Z. Bisgaard, S. S. Viftrup, L. Holmegaard, and Henrik Stapelfeldt, J. Chem. Phys. {\bf 125}, 194309 (2006).

\bibitem{Stapelfeldt-Seideman-RMP03} H. Stapelfeldt and T. Seideman, Rev. Mod. Phys. {\bf 75}, 543 (2003).

\bibitem{Seideman-JCP95} T. Seideman, J. Chem Phys. {\bf 103}, 7887 (1995).

\bibitem{Seideman-JCP99} T. Seideman, J. Chem Phys. {\bf 111}, 4397 (1999).

\bibitem{Seid-CPL96} T. Seideman, Chem. Phys. Lett. {\bf 253}, 279 (1996).

\bibitem{Seid-Hami-AAMOP05} T. Seideman and E. Hamilton, Advances In Atomic, Molecular, and Optical Physics {\bf 52}, 289-329 (2005).

\bibitem{Pero-Poul-PRL03} E. P\'eronne, M. D. Poulsen, C. Z. Bisgaard, H. Stapelfeldt, and T. Seideman, Phys. Rev. Lett. {\bf 91}, 043003 (2003).

\bibitem{Friedrich-Herschbach-PRL95} B. Friedrich and D. Herschbach, Phys. Rev. Lett. {\bf 74}, 4623 (1995).

\bibitem{Seideman-JCP01} T. Seideman, J. Chem Phys. {\bf 115}, 5965 (2001).

\bibitem{Poul-Pero-JCP04} M. D. Poulsen, E. P\'{e}ronne, H. Stapelfeldt, C. Z. Bisgaard, S. S. Viftrup, E. Hamilton and T. Seideman, J. Chem. Phys. {\bf 121}, 783-79 (2004).

\bibitem{Rena-Hert-PRA05} M. Renard, E. Hertz, S. Gu\'erin, H. R. Jauslin, B. Lavorel, and O. Faucher, Phys. Rev. A. {\bf 72}, 025401 (2005).

\bibitem{Leib-Aver-PRL03} M. Leibscher, I. Sh. Averbukh, and H. Rabitz, Phys. Rev. Lett. {\bf 90}, 213001 (2003).

\bibitem{Lee-Vill-PRL04} K. F. Lee, D. M. Villeneuve, P. B. Corkum, and E. A. Shapiro, Phys. Rev. Lett. {\bf 93}, 233601 (2004).

\bibitem{Zeng-Zhon-LaserPhys09} G. Zeng, F. Zhong, C. Wu, H. Jiang, and Q. Gong, Laser Physics {\bf 19}, pp.1691-1696.

\bibitem{Daem-Guer-PRL05} D. Daems, S. Gu\'erin, E. Hertz, H. R. Jauslin, B. Lavorel, and O. Faucher, Phys. Rev. Lett. {\bf 95},063005 (2005).

\bibitem{Ho-Mill-JCP09} P. J. Ho, M. R. Miller, and R. Santra, J. Chem. Phys. {\bf 130}, 154310 (2009).

\bibitem{Seid-PRL99} T. Seideman, Phys. Rev. Lett. {\bf 83}, 4971 (1999).

\bibitem{Pero-Poul-PRA04} E. P\'eronne, M. D. Poulsen, H. Stapelfeldt, C. Z. Bisgaard, E. Hamilton, and T. Seideman, Phys. Rev. A {\bf 70}, 063410 (2004).

\bibitem{Poul-Ejdr-PRA06} M. D. Poulsen, T. Ejdrup, H. Stapelfeldt, E. Hamilton and T. Seideman, Physical Review A {\bf 73}, 033405 (2006).

\bibitem{Hami-Seid-PRA05} E. Hamilton, T. Seideman, T. Ejdrup, M. D. Poulsen, C. Z. Bisgaard, S. S. Viftrup, H. Stapelfeldt, Phys. Rev. A {\bf 72},043402 (2005).

\bibitem{Bisg-Vift-PRA06} C. Z. Bisgaard, S. S. Viftrup, and H. Stapelfeldt, Phys. Rev. A {\bf 73},053410 (2006).

\bibitem{Bisg-Poul-PRL04} C. Z. Bisgaard, M. D. Poulsen, E. P\'eronne, S. S. Viftrup, H. Stapelfeldt, Phys. Rev. Lett {\bf 92}, 173004 (2004).

\bibitem{Guer-Rouz-PRA08} S. Gu\'{e}rin, A. Rouz\'{e}e, and E. Hertz,Phys. Rev. A {\bf 77}, 041404(R) (2008).

\bibitem{Kell-Dion-PRA00} A. Keller, C. M. Dion, and O. Atabek, Phys. Rev. A {\bf 61}, 023409 (2000).

\bibitem{Gers-Aver-PRA08} E. Gershnabel and I. Sh. Averbukh, Phys. Rev. A {\bf 78}, 063416 (2008).

\bibitem{Dion-Kell-PRA99} C. M. Dion, A. Keller, O. Atabek,and A. D. Bandrauk, Phys. Rev. A {\bf 59}, 1382 (1999).

\bibitem{Rouz-Guer-PRA06} A. Rouz\'{e}e, S. Gu\'{e}rin, V. Boudon, B. Lavorel, and O. Faucher, Phys. Rev. A {\bf 73}, 033418 (2006).

\bibitem{Ho-Sant-PRA08} P. J. Ho and R. Santra, Phys. Rev. A {\bf 78}, 053409 (2008).

\bibitem{Ho-Star-JCP09} P. J. Ho, D. Starodub, D. K. Saldin, V. L. Shneerson, A. Ourmazd, and R. Santra , J. Chem. Phys. {\bf 131}, 131101 (2009).

\bibitem{Buth-Sant-PRA08-1} C. Buth and R. Santra, Phys. Rev. A {\bf 77}, 013413 (2008).

\bibitem{Buth-Sant-JCP08} C. Buth and R. Santra, J. Chem. Phys. {\bf 129}, 134312 (2008).

%\bibitem{Santra-CP2006} R. Santra, Chem. Phy. {\bf 329}, 357 (2006).

\bibitem{Peterson-etal-APL08} E. R. Peterson, C. Buth, D. A. Arms, R. W. Dunford, E. P. Kanter, B. Kr\"assig, E. C. Landahl, S. T. Pratt, R. Santra, S. H. Southworth, and L. Young, Appl. Phys. Lett. {\bf 92}, 094106 (2008).

\bibitem{Arta-Seid-JCP08} M. Artamonov and T. Seideman, J. Chem Phys. {\bf 128}, 154313 (2008).

\bibitem{Larsen-etal-PRL00} J. J. Larsen, K. Hald, N. Bjerre, H. Stapelfeldt and T. Seideman, Phys. Rev. Lett. {\bf 85}, 2470 (2000).

\bibitem{Unde-Suss-PRL05} J. G. Underwood, B. J. Sussman, and A. Stolow, Phys. Rev. Lett. {\bf 94}, 143002 (2005).

\bibitem{Lee-Vill-PRL06} K. F. Lee, D. M. Villeneuve, P. B. Corkum, A. Stolow, and J. G. Underwood, Phys. Rev. Lett. {\bf 97}, 173001 (2006).

\bibitem{Vift-Kuma-PRA09} S. S. Viftrup, V. Kumarappan, L. Holmegaard, C. Z. Bisgaard, H. Stapelfeldt, M. Artamonov, E. Hamilton, and T. Seideman, Phys. Rev. A {\bf 79}, 023404 (2009).

\bibitem{Vift-Kuma-PRL07} S. S. Viftrup, V. Kumarappan, S. Trippel, H. Stapelfeldt, E. Hamilton, and T. Seideman, Phys. Rev. Lett. {\bf 99}, 143602 (2007).

\bibitem{Rouz-Guer-PRA08} A. Rouz\'{e}e, S. Gu\'{e}rin, O. Faucher, and B. Lavorel, Phys. Rev. A {\bf 77}, 043412 (2008).

\bibitem{Kim-Felker-JCP96} W. Kim and P. M. Felker, J. Chem. Phys. {\bf 104}, 1147 (1996).

\bibitem{Sakai-etal-JCP98} H. Sakai, C. P. Safvan, J. J. Larsen, K. M. Hilligs{\o}e, K. Hald, and H. Stapelfeldt, J. Chem. Phys. {\bf 110}, 10235 (1998).

\bibitem{Dick-Norr-PRL81} R. M. Dickson, D. J. Norris, W. E. Moerner, Phys. Rev. Lett. {\bf 81}, 5322 (1998).

\bibitem{Rouz-Rena-PRA07} A. Rouz\'{e}e, V. Renard, S. Gu\'{e}rin, O. Faucher, and B. Lavorel, Phys. Rev. A {\bf 75}, 013419 (2007).

\bibitem{Holm-Niel-PRL09} L. Holmegaard, J. H. Nielsen, I. Nevo, H. Stapelfeldt, F. Filsinger, J. K\"{u}pper, and G. Meijer, Phys. Rev. Lett. {\bf 102}, 023001 (2009).

\bibitem{Purc-Bark-PRL09} S. M. Purcell and P. F. Barker, Phys. Rev. Lett. {\bf 103},153001 (2009).

\bibitem{Larsen-etal-PRL99} J. J. Larsen, I. Wendt-Larsen, and H. Stapelfeldt, Phys. Rev. Lett. {\bf 83}, 1123 (1999).

\bibitem{Larsen-etal-JCP99} J. J. Larsen, H. Sakai, C. P. Safvan, I. Wendth-Larsen, and H. Stapelfeldt, J. Chem. Phys. {\bf 111}, 7774 (1999).

\bibitem{Even-Jort-JCP00} U. Even, J. Jortner, D. Noy, and N. Lavie, J. Chem. Phys. {\bf 112}, 8068 (2000).

%73
\bibitem{Drake-book} G. W. Drake, \textit{Atomic, Molecular, \& Optical Physics Handbook} (Amer Inst of Physics 1996), ISBN 978-1563962424.

\bibitem{Zare} R. N. Zare, \textit{Angular Momentum} (Wiley, New York, 1988).

\bibitem{Rohringer-Santra-PRA07} N. Rohringer and R. Santra, Phys. Rev. A {\bf 76}, 033416 (2007). 

\bibitem{Sakurai} J. J. Sakurai, \textit{Modern Quantum Mechanics} (Addison Wesley, Reading, 1993).

\bibitem{Kroto} H. W. Kroto, \textit{Molecular Rotation Spectra} (Wiley, London, 1975).

\bibitem{Reichl-book} L. E. Reichl, \textit{A Modern Course in Statistical Physics}, 3nd ed. (Wiley-VCH, Weinheim, 2009), ISBN 978-3527407828.

\bibitem{Blum-book} K. Blum, \textit{Density Matrix Theory and Applications, Physics of Atoms and Molecules}, 2nd ed. (Springer, 1996), ISBN 978-0306453410.

\bibitem{Loud-book} R. Loudon, \textit{The Quantum Theory of Light}, (Oxford University Press, Oxford, 1983).

\bibitem{Mand-Wolf-book} L. Mandel and E. Wolf, \textit{Optical Coherence and Quantum Optics}, (Cambridge University Press, 1995).

\bibitem{Land-Lifs-book1} L. D. Landau and  L. M. Lifshitz, \textit{Quantum Mechanics Non-Relativistic Theory, Third Edition: Volume 3} (Butterworth-Heinemann, 1981), ISBN 978-0750635394.

\bibitem{Paul-PR40} W. Pauli, Phys. Rev. {\bf 58}, 716-722 (1940).

\bibitem{Vars-Mosk-book} D. A. Varshalovich, A. N. Moskalev, and V. K. Khersonskii (Editor),\textit{Quantum Theory of Angular Momentum} (World Scientific Publishing Company, Singapore, 1988), ISBN 978-9971501075.

\bibitem{Als-Nielsen-McMorrow}  J. Als-Nielsen and D. McMorrow, \textit{Elements of Modern X-ray Physics} (Wiley, New York,  2001).

\bibitem{Mullges-book} G. Engeln-M\"ullges and F. Uhlig, \textit{Numerical Algorithms with Fortran} (Springer Verlag, Heidelberg, 1996) ISBN 3-540-60529-0.


\bibitem{Gerchberg-Saxton-Opt72} R. W. Gerchberg and W. 0. Saxton, Optik {\bf 35}, 237 (1972).

\bibitem{Saxton-Book78} W. 0. Saxton, \textit{Computer Techniques for Image Processing in Electron Microscopy} (Academic, New York, 1978).

\bibitem{Fienup-AO82} J. R. Fienup, Appl. Opt. {\bf  21}, 2758 (1982).

\bibitem{Fienup-etal-JOSA82} J. R. Fienup, T. R. Crimmins, and W. Holsztynski, J. Opt. Soc. Am. {\bf 72}, 610 (1982).

\bibitem{Fienup-JOSA87} J. R. Fienup, J. Opt. Soc. Am. A {\bf 4}, 118 (1987). 


\bibitem{Miao-Sayre-ACA00} J. Miao and D. Sayre, Acta Cryst. A {\bf 56}, 596 (2000). 

\bibitem{Marchesini-etal-PRB03} S. Marchesini, H. He, H. N. Chapman, S. P. Hau-Riege, A. Noy, M. R. Howells, U. Weierstall, and J. C. H. Spence, Phys. Rev. B {\bf 68}, 140101(R) (2003). 

\bibitem{Oszla-Suto-ACA04} G. Oszl\'anyic and A. S\"ut\H{o}, Acta Cryst. A {\bf 60}, 134 (2004).

\bibitem{Carrozzini-etal-ACA04} B. Carrozzini, G. L. Cascarano, L. De Caro, C. Giacovazzo, S. Marchesini, H. Chapman, H. He, M. Howells, J. S. Wu, U. Weierstall, and J. C. H. Spence, Acta Cryst. A{\bf60}, 331 (2004).

\bibitem{Luke-IP05} D. R. Luke, Inverse Probl. {\bf 21}, 37 (2005).

\bibitem {Wu-Spence-ACA05} J. S. Wu and J. C. H. Spence, Acta Cryst.  A {\bf 61}, 194 (2005).

\bibitem{Chapman-etal-JOSA06} H. N. Chapman, A. Barty, S. Marchesini, A. Noy, S. P. Hau-Riege, C. Cui, M. R. Howells, R. Rosen, H. He, J. C. H. Spence, U. Weierstall, T. Beetz, C. Jacobsen, and D. Shapiro, J. Opt. Soc. Am. A {\bf 23}, 1179 (2006).

\bibitem{Nugent-JOSA07} K. A. Nugent, J. Opt. Soc. Am. A {\bf 24} 536 (2007).

\bibitem{Marchesini-RSI07} S. Marchesini, Rev. Sci. Instrum. {\bf 78}, 011301 (2007). 

%\bibitem{Takahashia-etal-APL07}Y. Takahashia, Y. Nishino and T. Ishikawa and E. Matsubara, Appl. Phys. Lett. {\bf 90}, 184105 (2007).

\bibitem{Miao-etal-ARPC08} J. Miao, T. Ishikawa, Q. Shen, and T. Earnest, Annu. Rev. Phys. Chem. {\bf 59}, 387 (2008).

\bibitem{Robinson-etal-PRL01} I.K. Robinson, I.A. Vartanyants, G.J. Williams, M.A. Pfeifer, and J.A. Pitney, Phys. Rev. Lett. {\bf 87}, 195505 (2001).

\bibitem{Weierstall-etal-UM02} U. Weierstall, Q. Chen, J.C.H. Spence, M.R. Howells, M. Isaacson, and R.R. Panepucci, Ultramicroscopy {\bf 90}, 171 (2002).

\bibitem{Spence-etal-PTRSL02} J. C. H. Spence, U. Weierstall, and M. R. Howells, Philos. Trans. R. Soc. London {\bf 360}, 875 (2002).

\bibitem{Miao-etal-PRL02} J. Miao, T. Ishikawa, B. Johnson, E. H. Anderson, B. Lai, and K. O. Hodgson, Phys. Rev. Lett. {\bf 89}, 088303 (2002).

\bibitem{Nugent-etal-PRL03}  K. A. Nugent, A. G. Peele, H. N. Chapman, and A. P. Mancuso, Phys. Rev. Lett. {\bf 91}, 203902 (2003).

\bibitem{Miao-etal-PNAS03} J. Miao, K. O. Hodgson, T. Ishikawa, C.A. Larabell, M. A. LeGros, and Y. Nishino, Proc. Natl. Acad. Sci. U.S.A. {\bf 100}, 110 (2003).

\bibitem{Xiao-Shen-PRB05} X. Xiao and Q. Shen, Phys. Rev. B {\bf 72}, 033103 (2005).

\bibitem{Quiney-etal-NP06} H. M. Quiney, A. G. Peele, Z. Cai, D. Paterson, and K. A. Nugent, Nature Phys. {\bf 2}, 101 (2006).

\bibitem{Miao-etal-PRL06} J. Miao, C.-C. Chen, C. Song, Y. Nishino, Y. Kohmura, T. Ishikawa, D. Ramunno-Johnson, T.-K. Lee, and S. H. Risbud, Phys. Rev. Lett. {\bf 97}, 215503 (2006)

\bibitem{Joo-Taka-JMS02} D.-L. Joo, R. Takahashi, J. O'Reilly, H. Kat\^{o}, and M. Baba, Journal of Molecular Spectroscopy {\bf 215}, 155-159 (2002).

\bibitem{Hewe-Shen-JCP100} K. B. Hewett, M. Shen, C. L. Brummel, and L. A. Phillips, J. Chem. Phys. {\bf 100}, 4077 (1994).

\bibitem{Howa-Fall-MolPhys99} S. T. Howard, I. A. Fallism, and D. J. Willock, Molecular Physics {\bf  97}, 913-918 (1999). 

\bibitem{Smit-Ledi-RCMS98} D. J. Smith, K. W. D. Ledingham, R. P. Singhal, H. S. Kilic, T. McCanny, A. J. Langley, P. F. Taday, and C. Kosmidis, Rapid Communications in Mass Spectrometry {\bf 12}, 813-820 (1998).

\bibitem{Shen-Baza-JSyncRad04} Q. Shen, I. Bazarov and P. Thibault, J. Synchrotron Rad. {\bf 11}, 432-438 (2004).

\bibitem{Fils-Kuep-JCP09} F. Filsinger, J. K\"upper, G. Meijer, L. Holmegaard, J. H. Nielsen, I. Nevo, J. L. Hansen, and H. Stapelfeldt, J. Chem. Phys. {\bf 131}, 064309 (2009).

\bibitem{Stap-EuroPhysJD} H. Stapelfeldt, The European Physical Journal D {\bf 26}, 15 (2003).

\bibitem{Vill-Asey-PRL00} D. M. Villeneuve, S. A. Aseyev, P. Dietrich, M. Spanner, M. Y. Ivanov, and P. B. Corkum, Phys. Rev. Lett. {\bf 85}, 542--545 (2000).

\bibitem{Herzberg-book2} G. Herzberg, \textit{Molecular Spectra and Molecular Structure: II. Infrared and Raman Spectra of Polyatomic Molecules} (D. Van Nostrand Company (Canada) Inc., 1968), ISBN 978-0894642692.

\bibitem{Bunker-book} P. R. Bunker and P. Jensen, \textit{Molecular Symmetry and Spectroscopy 2nd edition} (National Research Council of Canada, 1998), ISBN 0-660-17519-3.

\bibitem{Herzberg-book3} G. Herzberg, \textit{Molecular Spectra and Molecular Structure: III. Electronic Spectra and Electronic Structure of Polyatomic Molecules} (D. Van Nostrand Company (Canada) Inc., 1966), ISBN 978-0894642708.

%\bibitem{Rose} M. E. Rose, \textit{Elementary Theory of Angular Momentum} (Wiley, New York, 1957).

\end{thebibliography}
\end{document}